\documentclass[12pt]{iopart}
\usepackage{iopams}  
\usepackage[numbers]{natbib}
\usepackage{amssymb,graphics,lineno}
\usepackage{rotating}
\usepackage{enumerate}
\usepackage{setspace}
\usepackage{multirow, color}
\usepackage[bookmarks=false]{hyperref}

\begin{document}

\newcommand{\stgb}[5]{\ensuremath{\Sigma#1\left(#2#3#4\right)\theta=#5^\circ}} 
\newcommand{\atgb}[8]{\ensuremath{\Sigma#1\left(#2#3#4\right)_1/\left(#5#6#7\right)_2\Phi=#8^\circ}} 
\newcommand{\planef}[3]{\ensuremath{\left\{ #1#2#3 \right\} }}
\newcommand{\plane}[3]{\ensuremath{\left( #1#2#3 \right)}}
\newcommand{\dirf}[3]{\ensuremath{\left\langle #1#2#3 \right\rangle}}
\newcommand{\dir}[3]{\ensuremath{\left[ #1#2#3 \right]}}

\title[Carbon Segregation to Fe Grain Boundaries]{Quantifying the Energetics and Length Scales of Carbon Segregation to $\alpha$-Fe Symmetric Tilt Grain Boundaries Using Atomistic Simulations}

\author{N.R. Rhodes$^{2}$, M.A. Tschopp$^{1,2,*}$, K.N. Solanki$^{3}$}
\address{$^1$ Oak Ridge Institute for Science and Education/Army Research Laboratory, Weapons and Materials Research Directorate, Aberdeen Proving Ground, MD 21005}
\address{$^2$ Center for Advanced Vehicular Systems, Mississippi State University, \\ Starkville, MS 39762}
\address{$^3$ School for Engineering of Matter, Transport, and Energy, Arizona State University, Tempe, AZ 85287}
\ead{mark.a.tschopp.ctr@mail.mil}

\begin{abstract}

Segregation of impurities to grain boundaries plays an important role in both the stability and macroscopic behavior of polycrystalline materials.  The research objective in this work is to better characterize the energetics and length scales involved with the process of solute and impurity segregation to grain boundaries. Molecular statics simulations are used to calculate the segregation energies for carbon within multiple substitutional and interstitial grain boundary sites over a database of 125 symmetric tilt grain boundaries in Fe.  The simulation results show that there are two energetically favorable grain boundary segregation processes: (1) an octahedral C atom in the lattice segregating to an interstitial grain boundary site and (2) an octahedral C atom and a vacancy in the lattice segregating to a grain boundary substitutional site.  In both cases, lower segregation energies than appear in the bulk lattice were calculated.  Moreover, based on segregation energies approaching bulk values, the length scale of interaction is larger for interstitial C than for substitutional C in the grain boundary ($\approx{5}$ \AA\ compared to $\approx{3}$ \AA\ from center of the grain boundary).  A subsequent data reduction and statistical representation of this dataset provides critical information such as about the mean segregation energy and the associated energy distributions for carbon atoms as a function of distance from the grain boundary, which quantitatively informs higher scale models with energetics and length scales necessary for capturing the segregation behavior of alloying elements and impurities in Fe.  The significance of this research is the development of a methodology capable of ascertaining segregation energies over a wide range of grain boundary character (typical of that observed in polycrystalline materials), which herein has been applied to carbon segregation to substitutional and interstitial sites in a specific class of grain boundaries in $\alpha$-Fe.


\end{abstract}

\submitto{\MSMSE}
\maketitle

\section{\label{sec:sec1}Introduction}

The computational design of future alloys will greatly depend on our ability to understand and quantify nanoscale phenomena in metallic material systems.  For instance, impurity segregation to grain boundaries (GBs) in alloys can have a profound effect on underlying microstructural processes, which can subsequently be detrimental to mechanical properties in polycrystals, e.g., hardness, toughness, and fracture behavior \cite{Pug1991,Lej1995,Sut1997,Hon1977,Hof1996,Bal1979,Bri1992,Foi1992,Hon1996,Sch2004,McM2004}. On the other hand, in some cases, atom segregation to GBs can actually be beneficial for macroscale material properties, e.g., by forming intermetallics, strengthening GB cohesion, or preventing grain growth \cite{McM2004,Che2004,Bub2006,Gen2000,Liu2004}. Segregation also plays a role in GB decohesion.  For instance, Yamaguchi et al.~recently showed that S segregation to Ni GBs leads to a reduction in GB tensile strength by an order of magnitude \cite{Yam2005}.  Moreover, Solanki et al.~found that certain H defects are favored at $\alpha$-Fe GBs and that these species affect the cohesive GB strength \cite{Sol2012}.  Since the presence of impurities and atoms at GBs can have such an acute impact on many material properties, understanding their interaction with and segregation to GBs and other lattice defects is crucial to the design of future materials.  

One potential application of work in atomic segregation is nuclear materials. Nuclear material design is also dependent upon understanding the segregation of impurities and defects within cladding materials. Radiation damage, through cascade events, ultimately results in numerous vacancies and interstitial atoms within the lattice. Impurities within the material then tend to diffuse with the vacancies or interstitial atoms as they attempt to return to equilibrium positions in the lattice \cite{Tak1993,Kam2001,Joh1976}. Such non-equilibrium radiation-induced segregation has a profound effect on material properties due to the accelerated segregation kinetics in comparison to the typical kinetics in thermal equilibrium \cite{Joh1976,Lam1978,Fau1996}. Moreover, since many cladding materials are polycrystalline and GBs are significant sinks for defect and impurity segregation, understanding impurity segregation to GBs is crucial to nuclear material design.

A number of studies have experimentally characterized the presence and effect of impurities on GBs in various materials \cite{Hon1977,Che2004,Bub2006,Tah2005,Mol1998,Sau2007,Kra1990,Lej2003,Lej1994,Kob2010,Tah2010,Ish2006,Set1999,Kra1998}. For instance, Lej\v{c}ek used Auger electron spectroscopy (AES) to show that segregants are equally distributed between fracture surfaces in symmetric tilt grain boundaries (STGBs) and distributed unevenly for asymmetric boundaries in Fe-Si bicrystals \cite{Lej1994}. Furthermore, Lej\v{c}ek et al.~comprehensively classified \dir100 tilt GBs in $\alpha$-Fe into special, vicinal, and general categories using AES measurements of GB segregation \cite{Lej2003}. Such studies have also proven useful in GB engineering.  Recently, Kobayashi et al.~used electron backscatter diffraction (EBSD) and orientation imaging microscopy (OIM) to show that intergranular embrittlement caused by sulfur segregation in nickel can be lessened by developing an optimal GB microstructure \cite{Kob2010}. Moreover, EBSD experiments of Al-Zr alloys have shown that GB sites in immobile twist GBs have a much higher degree of segregation than at mobile tilt GBs \cite{Tah2005}.  Researchers have also begun to use high-resolution transmission electron microscopy (TEM) and Local Electron Atom Probes (LEAP) \cite{Tah2010,Ish2006,Set1999,Kra1998} to create three-dimensional atom-by-atom representations of solute segregation at GBs and characterize their concentrations. For example, Taheri et al.~utilized a method that combined EBSD and focused ion beam milling specimen preparation with LEAP to measure solute segregation at GBs in an Al alloy \cite{Tah2010}. Furthermore, LEAP has been utilized by Isheim and colleagues to illustrate the reduction in impact toughness in low-carbon steels as a result of the combined segregation behavior of C, B, S, and P \cite{Ish2006}. While critical experiments provide valuable insight into solute segregation to GBs, techniques that aim to probe how atomic structure impacts segregation are often difficult to perform, expensive, and very time intensive.  Additionally, these sorts of experiments have yet to be used to study large numbers of boundaries with varying GB character, typical of real polycrystalline materials. 

Modeling and simulation of segregation to GBs at the atomic scale can also provide valuable insight into segregation processes in polycrystalline materials \cite{Sch2004,Yam2005,Ari1992,Mai1996,Zha2010,Liu2005,Chr2010,Yua2010,Wac2008,Che2006,Wu1996,Kar2008,Men1994,Mil2005,Sut1982,Olm2009,Hol2010,Wol1991,Lez2004,Kur2004,Gao2006,Mal2004}. Typically, modeling and simulation of GB segregation at the nanoscale uses ab initio simulations \cite{Sch2004,Yam2005,Ari1992,Mai1996,Zha2010,Liu2005,Chr2010,Yua2010,Wac2008,Che2006,Wu1996,Kar2008} or molecular dynamics (MD) \cite{Men1994,Mil2005,Sut1982,Olm2009,Hol2010,Wol1991,Lez2004,Kur2004,Gao2006,Mal2004}.  Ab initio calculations are often used to study the electronic effects of solute presence at GBs and their influence on cohesive strength.  For instance, Liu et al.~investigated the preferred site of Mg segregation at Al GBs and determined that Mg forms weaker metallic bonds with Al atoms in the GB region and decreases the cohesive strength of the GB \cite{Liu2005}.  Wachowicz and Kiejna \cite{Wac2008} studied the effect of substitutional and interstitial N, B, and O impurities at an Fe GB and found that N in both positions and interstitial B are embrittlers while O in both positions and substitutional B enhance GB cohesion. The segregation energies and cohesive effects of twenty impurities and alloying elements at a Zr twist GB were calculated by Christensen et al., who showed that most elements have an adverse effect on GB cohesion, with Cs being the most embrittling \cite{Chr2010}. These techniques, however, can be computationally expensive and have typically been used only for a few GBs.   On the other hand, MD studies often use empirical or semi-empirical interatomic potentials fit to ab initio and experimental properties.  These simulations are much less expensive than their ab initio counterparts but are limited by the accuracy or availability of interatomic potentials.  Nonetheless, MD simulations are increasingly being used to study GB segregation in both fcc and bcc materials. Millett et al.~investigated the impact of dopants at a Cu GB and concluded that, for a particular concentration of each dopant atomic size, the thermodynamic driving force for grain growth could be eliminated \cite{Mil2005}. Lezzar et al.~concluded that the driving force for intergranular segregation in Ag(Ni) and Ni(Ag) systems can be primarily attributed to the atomic size effect \cite{Lez2004}. While MD has been more commonly used for fcc materials, such simulations have also provided insight into GB segregation in body-centered cubic (bcc) Fe as well \cite{Kur2004,Gao2006,Mal2004}. For instance, Gao et al.~used MD simulations to show that, at $\alpha$-Fe GBs, He binding energy increases with excess volume and binds to GBs more strongly in interstitial positions than in substitutional ones \cite{Gao2006}. Additionally, Malerba et al.~modeled displacement cascades in an Fe-Cr system with MD to show that a large percentage of Cr atoms are located in interstitial clusters, which may greatly reduce the mobility of interstitial loops when compared to pure Fe \cite{Mal2004}.

While MD simulations are much less expensive than ab initio simulations, very few simulations consider a large number of GBs in their analysis of GB-related properties. GBs have five degrees of freedom associated with them (plus three associated with translation at an atomistic level), and many experimental methods have begun to measure the GB character in terms of these degrees of freedom \cite{Say2002,Kim2005,Say2004} for GB engineering purposes.  However, in nanoscale calculations, only a few studies have explored fifty or more GBs in their analysis of nanoscale properties.  Tschopp and McDowell have shown that asymmetric tilt GB systems in Cu and Al facet into the structural units of their corresponding symmetric tilt GB counterparts \cite{Tsc2007b, Tsc2007c, Tsc2007} and that the grain boundary structure results in very different dislocation nucleation properties and mechanisms \cite{Spe2007, Tsc2008, Tsc2008a, Tsc2008c}. Holm et al.~calculated energies of 388 GBs in Al, Au, Cu, and Ni, and observed that the GB energy scales with the shear modulus and that boundaries with significant stacking fault character correlate with the stacking fault energy \cite{Hol2010}. The classic work of Wolf has shown that, for several Mo and Fe GB systems, GB energy correlates nearly linearly with volume expansion per unit area \cite{Wol1991}. The recent work of Tschopp et al.~used $>$150 Fe STGBs to demonstrate that, based on formation energies, self-interstitial atoms display a larger energetic driving force for binding to GBs than vacancies do \cite{Tsc2011, Tsc2012}. Clearly, a similar methodology using molecular statics and dynamics simulations that can analyze how segregation in $\alpha$-Fe systems is influenced by GB character would be valuable to understanding GB segregation and, perhaps, to engineering materials by increasing beneficial GBs while decreasing detrimental GBs.

In this work, the research objective is to quantify the energetics and length scales associated with C segregation to Fe GBs.  The Fe-C system is chosen as an ideal system because C is known both experimentally \cite{Cow1998,Pap1971,Tak2005} and computationally \cite{Hon1989,Yan2004,Rud2009,Vei2010,Hri2011} to segregate to the boundaries, dislocations, and surfaces; moreover, a number of experimental studies have shown that C increases the GB cohesive strength in $\alpha$-Fe \cite{Cow1998,Pap1971,Tak2005}.  The methodology used here provides a means for simulating how GB character impacts the segregation of C to a large number of Fe GBs.  In this work, we utilize an interatomic potential \cite{Hep2008} specifically formulated to capture the energetics of C interactions with point defects in $\alpha$-Fe, which is in good agreement with \textit{ab initio} results \cite{Dom2004}.  This paper is outlined as follows.  Section \ref{sec:sec2} describes the simulation methodology used to simulate and calculate segregation data.  Section \ref{sec:sec3} discusses the results of the simulations and their significance for modeling GB segregation.  Section \ref{sec:sec4} discusses our results, particularly addressing how the present methodology may need to be extended to better model the interaction between GBs and interstitial atom species.  Section \ref{sec:sec5} summarizes this research and provides conclusions based on our results.

\section{\label{sec:sec2}Simulation Methodology}

In this work, the segregation energy associated with a single C atom was calculated at sites within or around $\alpha$-Fe GBs.  While it is well known that C occupies octahedral interstitial sites in the perfect single crystal $\alpha$-Fe lattice, here we examined a few different scenarios for C.  As a first order approximation, we utilize sites formed on the initial GB lattice.  This process is meant to mimic the restructuring at the boundary that occurs through an interaction with a vacancy and then a subsequent occupation of an available site in the restructured boundary (at the exact location of the initial vacancy).  The hypothesis is that at a GB that has undergone some restructuring due to interactions with point defects, such segregation processes and sites may be energetically favorable.  This may be a reasonable assumption given that DFT calculations have shown that another interstitial atom, N, has very similar formation energies in both substitutional and interstitial sites within a $\Sigma5$\plane210 GB \cite{Wac2008}.  Moreover, in an ultra-low C bake-hardening steel sheet, three-dimensional atom probe measurements have found that C atom concentrations at the GB can be more than 200 times that in the bulk \cite{Tak2005}; hence the possibility for C to segregate to both interstitial and substitutional sites is probable.  In Section \ref{sec:sec4}, this assumption will be further compared with interstitial sites within the GB lattice (as opposed to sites directly on top of the GB lattice) for a few GBs, based on starting coordinates obtained from a Voronoi tesselation of the simulation cell.   The first-order process used to calculate the segregation energies of C in $\alpha$-Fe is as follows:
\begin{enumerate}
	\item A GB is selected from a GB database that contains 125 STGBs (50 \dirf100, 50 \dirf110, 25 \dirf111).
	\item A GB site (within 15 \AA) is chosen and a C atom is substituted for the Fe atom at this site.
	\item A molecular dynamics code (LAMMPS \cite{Pli1995}) is used to minimize the energy of the GB with the substitutional C atom.
	\item The GB, site position, and calculated segregation energy of the substitutional C atom are stored.
	\item The process is repeated for all sites within 15 \AA\ of the GB center and for all GBs within the GB database.
\end{enumerate}

The Hepburn and Ackland Fe-C interatomic potential \cite{Hep2008} is used to model the Fe GBs and their interaction with the subsitutional C atom. This potential is based on the embedded-atom method (EAM) formalism \cite{Daw1983,Daw1984} and is in agreement with Density Functional Theory with respect to the energetics pertaining to interactions between C atoms and Fe self-interstitial atoms, vacancies, and other C atoms. Unlike prior Fe-C potentials, the Hepburn-Ackland Fe-C potential was the first EAM potential to correctly capture covalent bonding of two C atoms within a vacancy.  Moreover, previous EAM potentials showed strong binding of C to overcoordinated defects, such as self-interstitial atoms, whereas the Hepburn-Ackland potential correctly captures the strong repulsion between orvercoordinated defects and C, in agreement with ab initio results.  This repulsion can be important for interactions between C atoms and GBs.  Last, this potential has been simulated at temperature yielding dynamics and mechanisms\footnote{The migration energy, $E_m=0.887$ eV, is in good agreement with the \textit{ab initio} results of Domain et al.~\cite{Dom2004}, $E_m=0.902$ eV.  Additional dynamic simulations by Hepburn and Ackland \cite{Hep2008} at 1400 K with a single C atom in an Fe single crystal lattice show that C exclusively migrates from octahedral to octahedral sites through the tetrahedral sites.} in agreement with \textit{ab initio} results \cite{Dom2004}.  For instance, Terentyev et al.~\cite{Ter2011} recently used this potential to investigate the influence of C atoms on the stability and migration of small clusters of point defects and found that C atoms have an attractive interaction with vacancy clusters containing fewer than four vacancies.  This potential provides a reasonably accurate representation of the Fe-C system and is deemed appropriate for studies of single C atoms within the bcc Fe lattice.

The segregation energy is calculated for C as a function of position at each site within 15 \AA\ of the GB. For each GB structure, an Fe atom at a particular site $\alpha$ is replaced with a C atom and the simulation cell is relaxed using the Polak-Ribi\'{e}re conjugate gradient energy minimization process.  The total energy of the simulation cell is calculated and the process is repeated for each atomic site within each GB in the database. The segregation energy calculations follow a similar approach to others, e.g., Liu et al.~\cite{Liu2005}. The segregation energy associated with a C atom at site $\alpha$ is calculated with

\begin{equation}
  \label{eq:eq1}
	E_{seg}^{C^\alpha}=\left(E_{tot}^{GB,C^\alpha_{sub}}- E_{tot}^{GB}\right)-\left(E_{bulk}^{Fe,C_{sub}}- E_{bulk}^{Fe}\right)	
\end{equation}

\noindent where $E_{tot}^{GB,C^\alpha_{sub}}$ and $E_{tot}^{GB}$ are the total energies of the GB structure with and without the solute substitution.  $E_{bulk}^{Fe,C_{sub}}$ and $E_{bulk}^{Fe}$ are the total energies of a single crystal bulk Fe simulation cell with and without the substituted C solute. The bulk energies used a 10$a_0$x10$a_0$x10$a_0$ bcc cell with 2000 atoms.  Hence, $E_{bulk}^{Fe}$ is equal to $2000E_{c}^{Fe}$ ($E_{c}^{Fe}=4.013$ eV, i.e., cohesive energy of Fe) and $E_{bulk}^{Fe,C_{sub}}=2000E_{c}^{Fe}+0.391$ eV. These bulk energies are subtracted in Eq.~\ref{eq:eq1} to remove the effect of substituting the C atom.  Hence, $E_{seg}^{C^\alpha}\approx0$ represents that substituting a C atom into site $\alpha$ in the GB simulation cell results in an equivalent energy difference as substituting the C atom into a perfect bcc Fe lattice.  As with prior work, a negative value of $E_{seg}^{C^\alpha}$ represents that it is energetically favorable for C to bind to site $\alpha$ compared to the bulk lattice.  It should be noted that the segregation energy, $E_{seg}^{C^\alpha}$, as defined is equal to $(-1)$ times the binding energy $E_{b}^{C\alpha}$.  Using the same terms as in Eq.~\ref{eq:eq1}, the binding energy is typically defined as

\begin{equation}
  \label{eq:eq2}
	E_{b}^{C^\alpha}=\left(E_{tot}^{GB}+E_{bulk}^{Fe,C_{sub}} \right)-\left(E_{tot}^{GB,C^\alpha_{sub}}+E_{bulk}^{Fe}\right)=-E_{seg}^{C^\alpha}.
\end{equation}

\noindent The segregation energy in Eq.~\ref{eq:eq1} will be used for the subsequent analysis of a C atom in substitutional GB sites.  Later, in Section \ref{sec:sec4}, the segregation energy equation is discussed in the context of two different scenarios: (1) a C atom in an octahedral site and a vacancy in the bulk lattice combining into C at a GB substitutional site and (2) a C atom at an octahedral site occupying GB interstitial sites.  That is, the present definition of segregation energy in Eq.~\ref{eq:eq1} is modified for the Fe-C system to account for the fact that C does not occupy substitutional sites within the bulk lattice.  This energy difference is $0.930$ eV; hence, $E_{seg}^{C^\alpha} < -0.930$ eV is necessary for Scenario 1 (octahedral C and vacancy to GB substitutional C) to be energetically favorable. The method outlined in this section was used for each site in all 50 \dirf100 STGB, as well as 50 \dirf110 and 25 \dirf111 STGBs. For each GB, the segregation energies were calculated as a function of atomic location.

\section{\label{sec:sec3}Simulation Results}

\subsection{Grain Boundary Structure and Energy}

The GB structure database used in the simulations herein contained 50 \dirf100, 50 \dirf110, and 25 \dirf111 STGBs. Bicrystal simulation cells with three-dimensional periodic boundary conditions were used to create the database \cite{Rit1996,Tsc2007,Tsc2007b,Tsc2007c}. To remove any possible interaction between the two boundaries, a minimum distance of 12 nm was used between them during generation. As with past work \cite{Tsc2007,Tsc2007b,Tsc2007c}, an atom deletion criterion along with multiple initial configurations with various in-plane rigid body translations were utilized to accurately obtain optimal minimum energy GB structure via the nonlinear conjugate gradient energy minimization process.

\begin{figure}[t!]
  \centering
	\includegraphics[width=3.5in,angle=0]{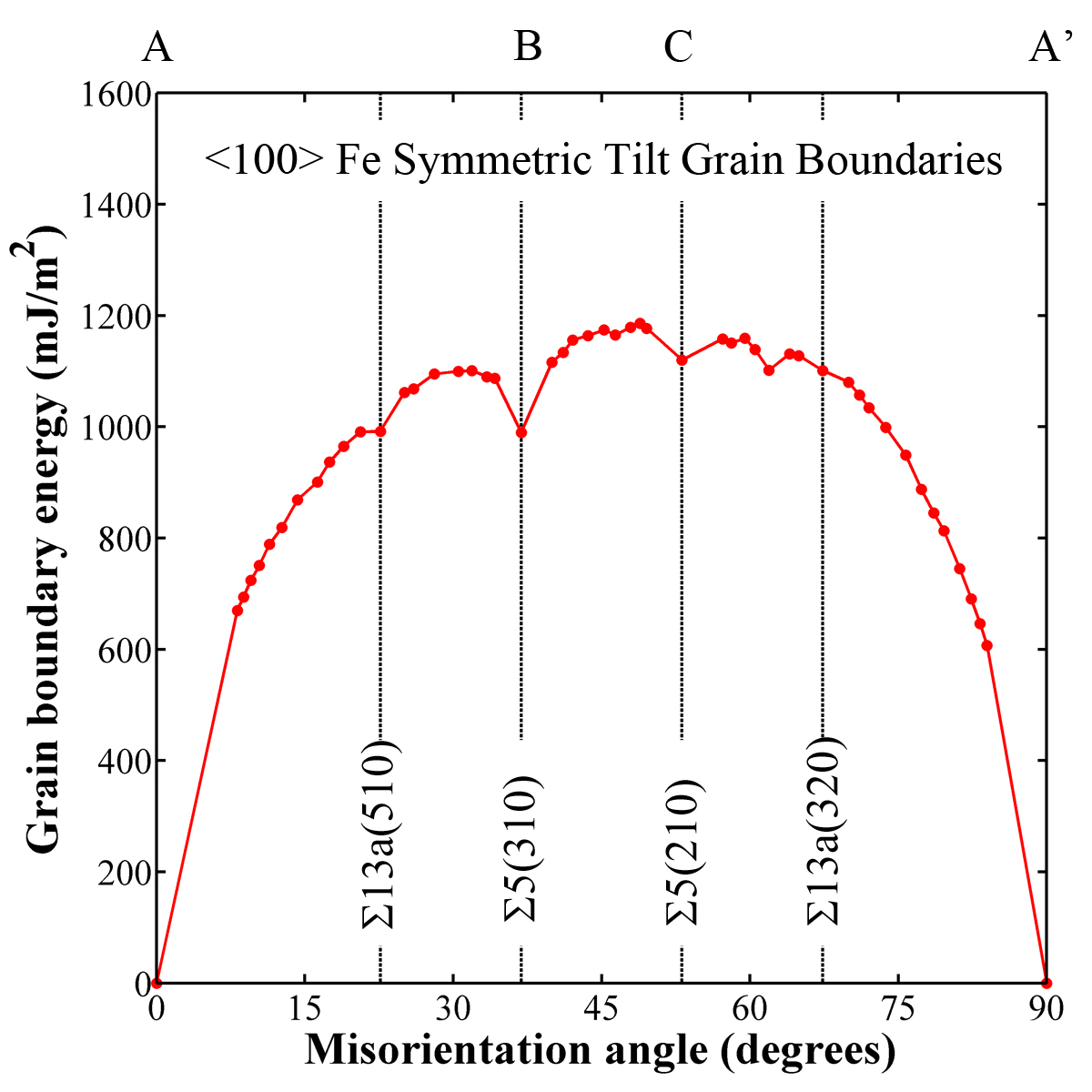} 
\caption{ \label{fig:fig1} \dirf100 symmetric tilt grain boundary energy as a function of misorientation angle \cite{Tsc2012}. The low-$\Sigma$ grain boundaries ($\Sigma\leq13$) in each system are identified. }
\end{figure}

The structures and energies of STGBs may be important to understand the interaction between C atoms and the boundary.  To examine the range of GB structures and energies that might be seen in polycrystalline materials, different grain boundaries from several GB tilt systems were used in the present simulations. The database used in this work is an expanded version of that first utilized in Tschopp et al.~\cite{Tsc2011}.  The \dirf100, \dirf110, and \dirf111 STGB systems chosen have several low order coincident site lattice (CSL) grain boundaries (e.g., $\Sigma$3, $\Sigma$5, $\Sigma$9, $\Sigma$11, and $\Sigma$13 boundaries), as well as both general high angle boundaries and low angle grain boundaries ($\theta\le15^\circ$). The GB energy as a function of misorientation angle for the \dirf100 STGB system is shown in Fig.~\ref{fig:fig1}.  This plot is similar to that found previously in Fe-Cr simulations \cite{Shi2008} and similar to misorientation-energy relationships found in fcc metals \cite{Rit1996,Wol1989,Wol1989a,Wol1990,Wol1990a}. The low-order CSL grain boundaries for the \dirf100 STGB system ($\Sigma$5 and $\Sigma$13 boundaries) are also illustrated in this figure. For the \dirf100 tilt axis, only minor cusps were observed in the energy relationship, most noticeably at the $\Sigma$5\planef310 boundary (990 mJ/m$^2$). In addition to many general high angle boundaries, several low angle boundaries ($\theta\le15^\circ$) are also plotted. The range of GB energies sampled was ~500 mJ/m$^{2}$.

\begin{figure}[t!]
  \centering
	\includegraphics[width=\textwidth,angle=0]{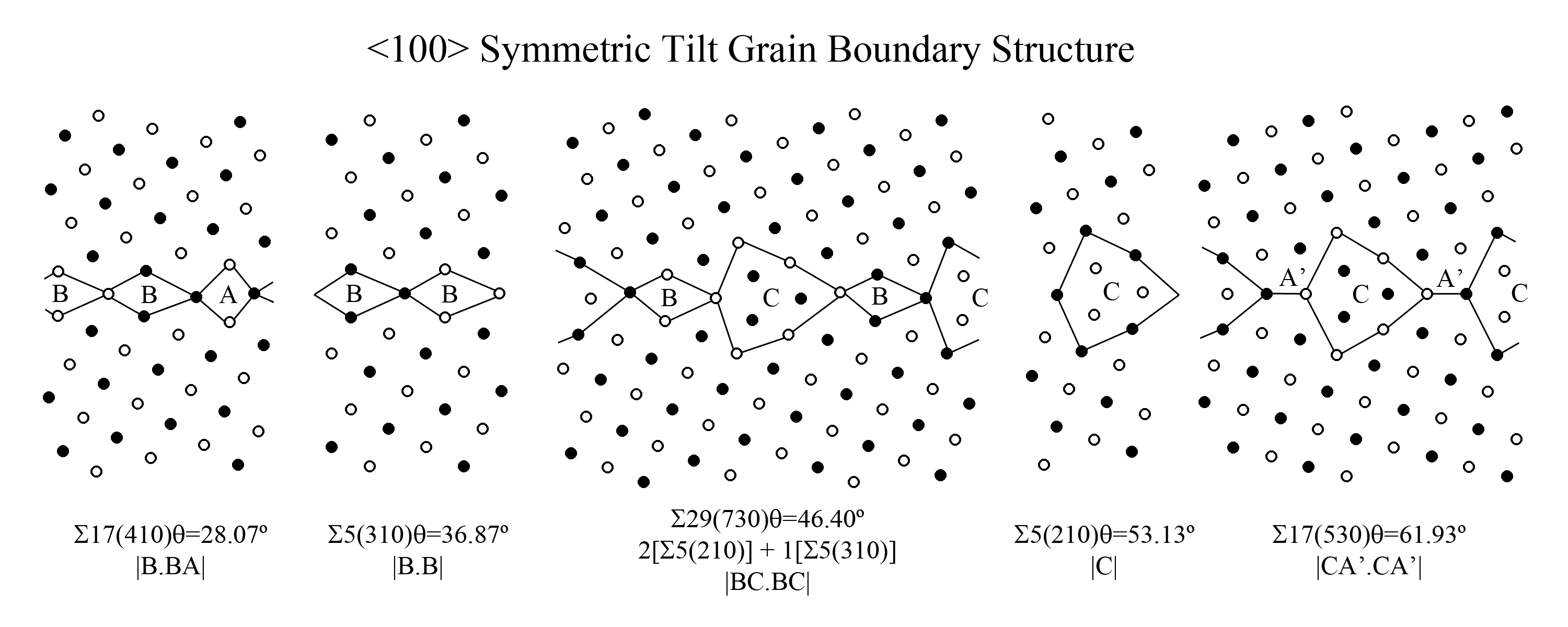} 
\caption{ \label{fig:fig2} \dirf100 symmetric tilt grain boundary structures with structural units outlined for the $\Sigma17$\plane410, $\Sigma5$\plane210, $\Sigma29$\plane730, $\Sigma5$\plane310, and $\Sigma17$\plane530 boundaries \cite{Tsc2012}.  Black and white denote atoms on different \planef100 planes. The different structural units are labeled A, B, C, and A'. }
\end{figure}

The GB structure plays an important role on the GB properties \cite{Mis2010}. For low angle boundaries, the GB is composed of an array of discrete dislocations and the corresponding energy can be calculated based on the classic Read-Shockley dislocation model.  However, at higher misorientation angles, the spacing between dislocations is small enough that dislocation cores overlap and dislocations rearrange to minimize the energy of the boundary. The resulting GB structures are often characterized by structural units \cite{Sut1983}. Grain boundaries with certain misorientation angles (and typically a low $\Sigma$ value) correspond to ``favored'' structural units, while all other boundaries are characterized by structural units from the two neighboring favored boundaries. An example of structural units in the \dirf100 STGB system is shown in Fig.~\ref{fig:fig2}, where the two $\Sigma$5 boundaries are favored STGBs, and the $\Sigma$29(730) boundary is a combination of structural units from the two $\Sigma$5 boundaries. The structural units for the $\Sigma$5(210) and $\Sigma$5(310) STGBs are labeled B and C, respectively, in a convention similar to that used for face-centered cubic metals \cite{Rit1996}. Also, notice that the ratio of structural units in the $\Sigma29$ GB can be determined by the crystallographic relationship of the two favored boundaries, \textit{i.e.},  $\Sigma29\plane730=2\left[\Sigma5\plane210\right]+1\left[\Sigma5\plane310\right]$.  In a similar manner, the two $\Sigma17$ boundaries are combinations of the favored B and C structural units and ``structural units'' of the perfect lattice, A and A'.

\subsection{\texorpdfstring{Segregation Energy for \dirf100 Boundaries}{Segregation Energy for 100 Boundaries}}

\begin{figure}[t!]
  \centering
	\includegraphics[width=\textwidth,angle=0]{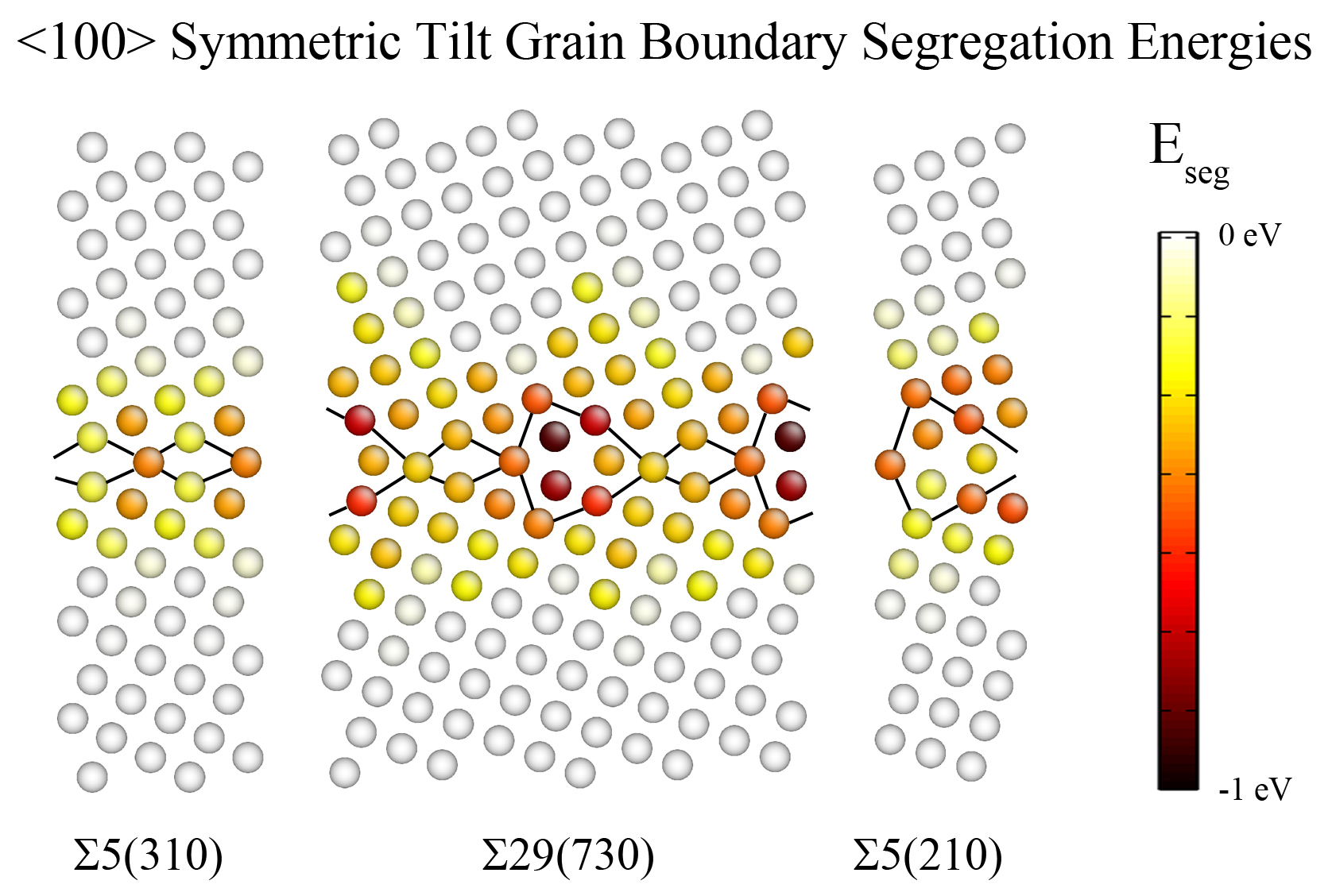} 
\caption{ \label{fig:fig3} Segregation energy as a function of site location for substitutional C atom in the $\Sigma5$\plane210, $\Sigma29$\plane730, and $\Sigma5$\plane310 boundaries.  }
\end{figure}

The segregation energies that correspond to the atomic positions in the middle three GB structures (Fig.~\ref{fig:fig2}) are shown in Figure \ref{fig:fig3}.  AtomEye is used to visualize the simulation results \cite{Li2003}.  In this graph, the color bar is normalized by subtracting the energy of substitutional C in the bulk so that the difference in energy between sites near the GB and in the bulk can be easily compared (i.e., atoms colored white have bulk segregation energies).  For all three GBs, the segregation energy becomes lower as the sites are located closer to the GB, meaning that segregation to the GB is favored for substitutional C.  However, there is not a simple gradient of the segregation energy from the GB center; the local structure also plays a pivotal role in the segregation energy.  For sites located farther from the GB, the segregation energy approaches that of the bulk, as denoted by segregation energies close to 0 eV.  Interestingly, although the structural units are the same between these three grain boundaries, there are some segregation energies in the $\Sigma29$\planef730 that are lower than either of the favored $\Sigma$5(210) and $\Sigma$5(310) STGBs, e.g., inside the C structural unit.  That is, the elastic interaction between differing structural units may produce a different distribution of segregation energies than a boundary composed of all the same structural unit.  While these trends seem to indicate a driving force for the segregation of C atoms from the bulk to the GB, this segregation energy needs to be further augmented ($E_{seg}^{C^\alpha}+0.930$ eV) to account for the fact that C lies in octahedral interstitial sites in the bulk, as will be further discussed in Section \ref{sec:sec4}. 

\begin{figure}[t!]
  \centering
	\begin{tabular}{cc}
	\includegraphics[width=2.9in,angle=0]{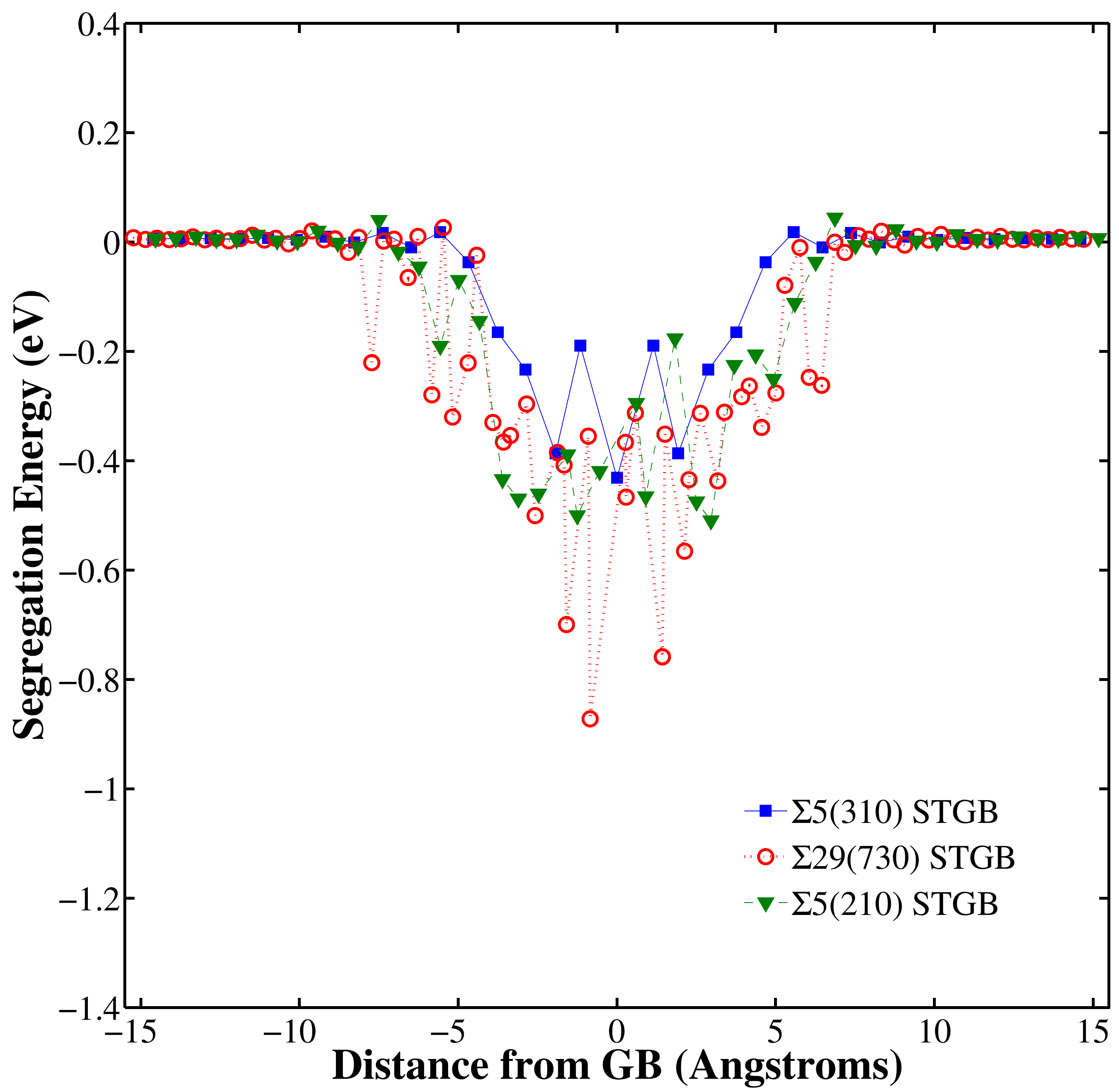} &
	\includegraphics[trim = 0.5in 0.0in 0.0in 0.75in, clip, width=2.9in,angle=0]{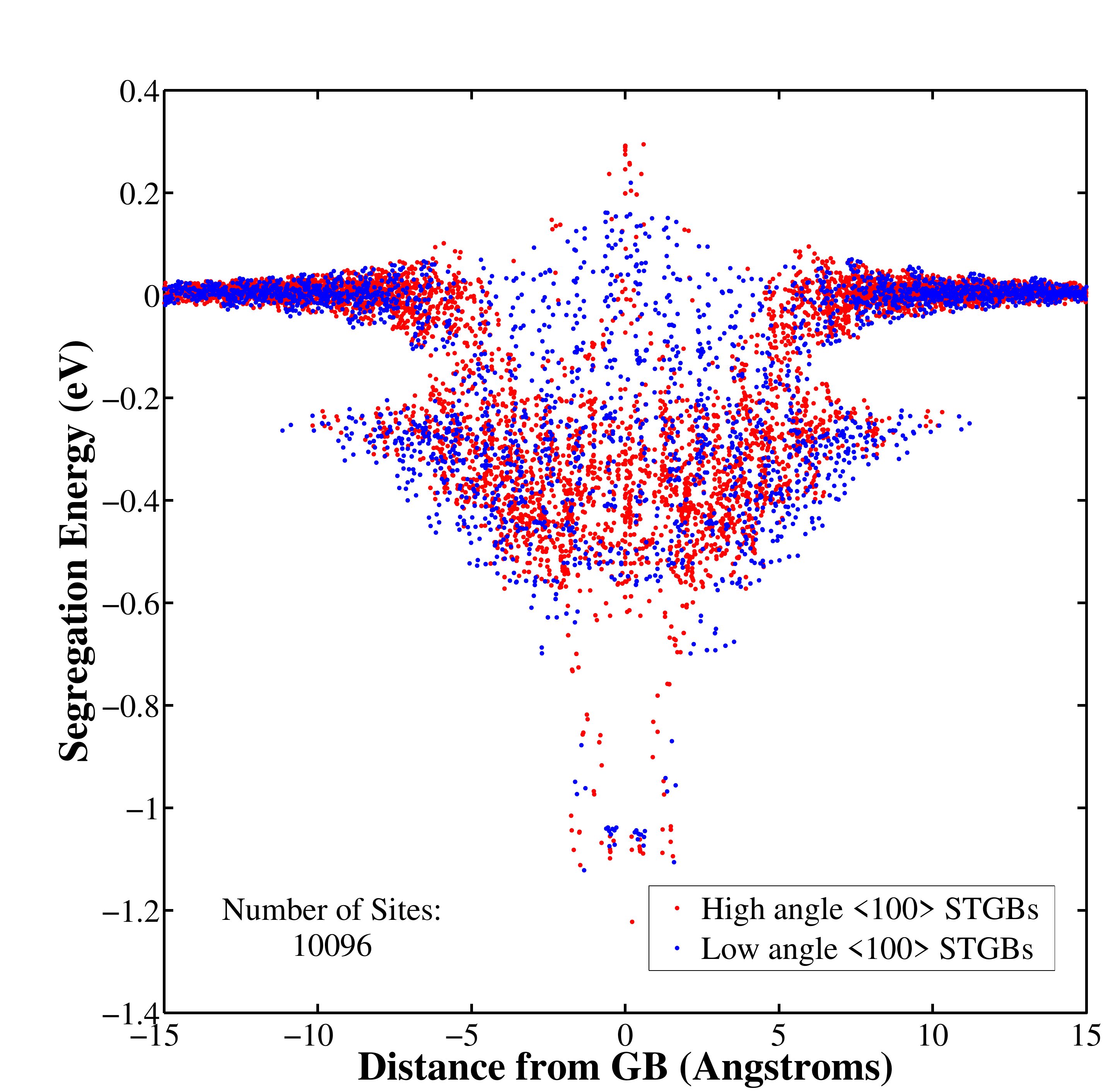} \\
	\textbf{(a) }& \textbf{(b)}
	\end{tabular}
\caption{ \label{fig:fig4} Distribution of segregation energies as a function of distance from the grain boundary for (a) the $\Sigma5$\plane210, $\Sigma29$\plane730, and $\Sigma5$\plane310 grain boundaries and (b) all 50 \dirf100 STGBs.  $E_{seg}^{C^\alpha} < 0.0$ eV indicates it is energetically favorable for substitutional C to bind to a substitutional site at the grain boundary.  However, since C occupies octahedral sites in the bulk lattice, $E_{seg}^{C^\alpha} < -0.930$ eV is required for an octahedral C atom and a vacancy to combine and bind to a GB substitutional site.}
\end{figure}

Plotting segregation energy against distance from the GB shows information similar to that in Figure \ref{fig:fig3}, but provides a convenient method to display the segregation energies of the sites in many different GBs at once.  The distribution of segregation energies as a function of distance for the three GB structures seen in Figure \ref{fig:fig3} is shown in Figure \ref{fig:fig4}.  Near the GB, all three GBs show a trend of negative segregation energies at sites near the boundary, which is the same behavior reflected in Figure \ref{fig:fig3}.  Moreover, notice the lack of any segregation energies that are near bulk values within 5 \AA\ of the GB center for these three boundaries.  Figure \ref{fig:fig4}b is a plot of the same distribution for all 50 \dirf100 STGBs, which includes both low angle ($\theta\le15^\circ$) and high angle grain boundaries.  As noted in Figure \ref{fig:fig4}b, over $10,000$ simulation sites (and atomistic simulations) were considered herein.  Most of the segregation energies that differ from that of the bulk occur between the GB center and about 7 \AA. While the majority of sites within this region have segregation energies less than that of the bulk, there are also a few GB sites that have segregation energies that are higher than in the bulk; most of these sites tend to be located along the centerline of the boundary.  There are a cluster of sites around 7-12 \AA\ from the GB that have segregation energies lower than the bulk as well.  There is a subtle difference between low and high angle boundaries.  Within 5 \AA\ of the GB center, low angle grain boundaries tend to have some segregation energies that are similar to the bulk values.  This is as expected, though.  Low angle boundaries are composed of dislocations separated by regions of perfect single crystal, which have similar segregation energies to bulk energies.  

\begin{figure}[t!]
  \centering
	\begin{tabular}{cc}
	\includegraphics[width=2.9in,angle=0]{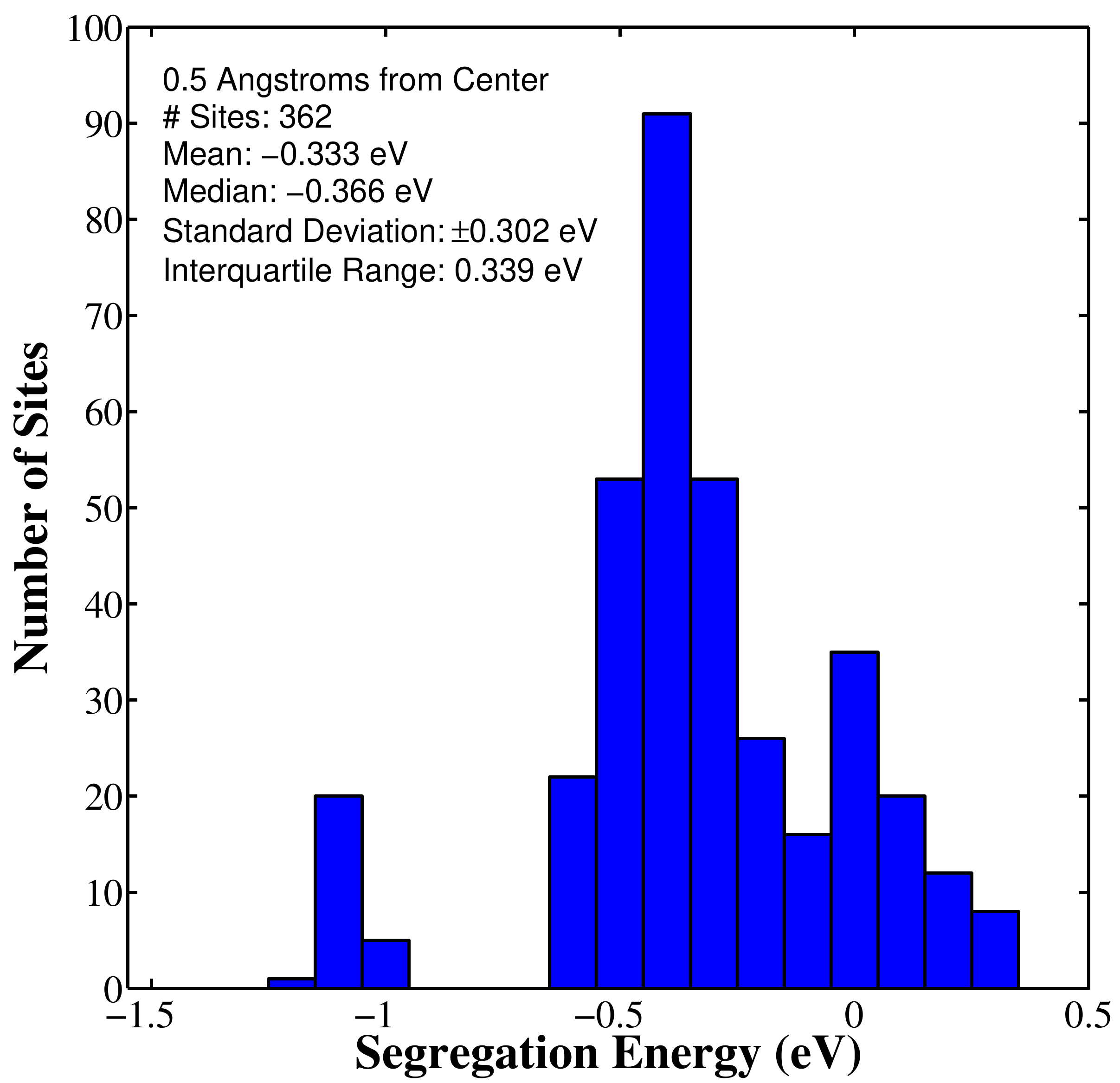} & 
	\includegraphics[width=2.9in,angle=0]{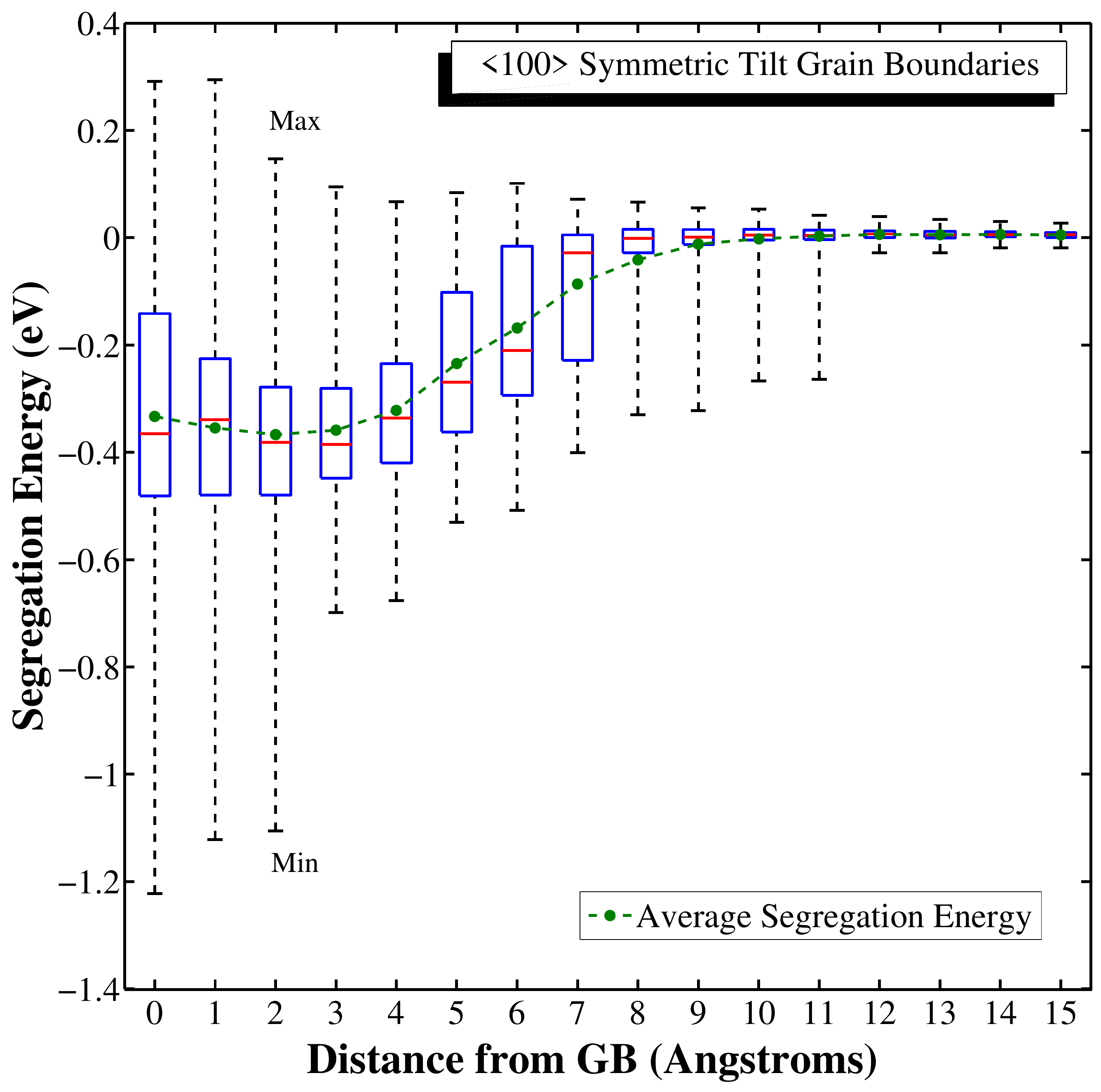} \\
	\textbf{(a) }& \textbf{(b)}
	\end{tabular}
\caption{ \label{fig:fig5} (a) The distribution of segregation energies within 0.5 \AA\ of the grain boundary center and the associated statistics.  (b) Boxplots of segregation energy as a function of distance from the grain boundary for all 50 \dirf100 STGBs.  The data is divided into 1 \AA\ bins, and a boxplot is made for each bin. The red lines are medians, the blue box ends are the first and third quartiles, and the black whisker ends are minimum and maximum values.  The mean segregation energy is also plotted in green.  }
\end{figure}

One way to represent the segregation energies-distance relationship is to bin the energies according to their distance from the GB center and to analyze the statistics associated with each bin (Figure \ref{fig:fig5}).  Due to the symmetric nature of the GB segregation energies as a function of distance (Figure \ref{fig:fig4}), the absolute value of the distance from the GB center was used to provide more data points for the statistical analysis.  Furthermore, the energies are split into 1 \AA\ bins to characterize the distributions and compute statistics for sites at a given distance from the GB.  An example of the 0 \AA\ bin ($-0.5$ \AA\ to $+0.5$ \AA) is shown in Figure \ref{fig:fig5}a along with several statistics: \# of boundaries, mean, median, standard deviation, and interquartile range\footnote{The interquartile range is defined as the difference between the 25\% percentile and 75\% percentile.}.  Once the appropriate statistics are calculated, a boxplot (Figure \ref{fig:fig5}b) is used to represent the segregation energy statistics in each bin, i.e., the minimum, 25\% percentile, median, 75\% percentile, and maximum segregation energies.  In the boxplot, the red line in the box is the median while the top and bottom edges of the blue boxes represent the 25\% and 75\% quartiles. The whiskers extending from the boxes cover the remainder of the range of energies for each bin, and the ends of the whiskers denote the maximum and minimum values of the segregation energies for each bin.  The mean value of the segregation energies in each bin is also plotted in green.  Boxplots can be very useful for displaying asymmetric distributions.

The mean segregation energy is lowest with sites close to the GB, as shown in Figure \ref{fig:fig5}, and it approaches the normalized bulk value of zero as sites are located farther from the boundary. Interestingly, the lowest mean segregation energies actually occur a few Angstroms from the center of the boundary.  Furthermore, at approximately $>$8 \AA, the boxes are closely centered about the bulk value, which shows that the overwhelming majority of atomic sites display a segregation energy similar to the bulk value.  However, it is noticed that there are a number of sites with segregation energies significantly below the bulk value that still persist up to approximately 11 \AA.  This trend indicates that it may be energetically favorable for substitutional C to segregate to sites within 11 \AA\ of the GB, albeit there is a much larger driving force with decreasing distance from the boundary. Additionally, the majority of bins display energy distributions that are skewed, usually in the direction of negative energy, i.e., the median is closer to the lower edge of the box (mainly for distances less than 7 \AA).  While the median fluctuates somewhat, the mean segregation energies - which track with the median - follow a much smoother relationship with distance.

\subsection{\texorpdfstring{Segregation Energy for \dirf110 and \dirf111 Boundaries}{Segregation Energy for 110 and 111 Boundaries}}

\begin{figure}[t!]
  \centering
	\includegraphics[trim = 0.5in 0.0in 0.0in 0.75in, clip, width=4in,angle=0]{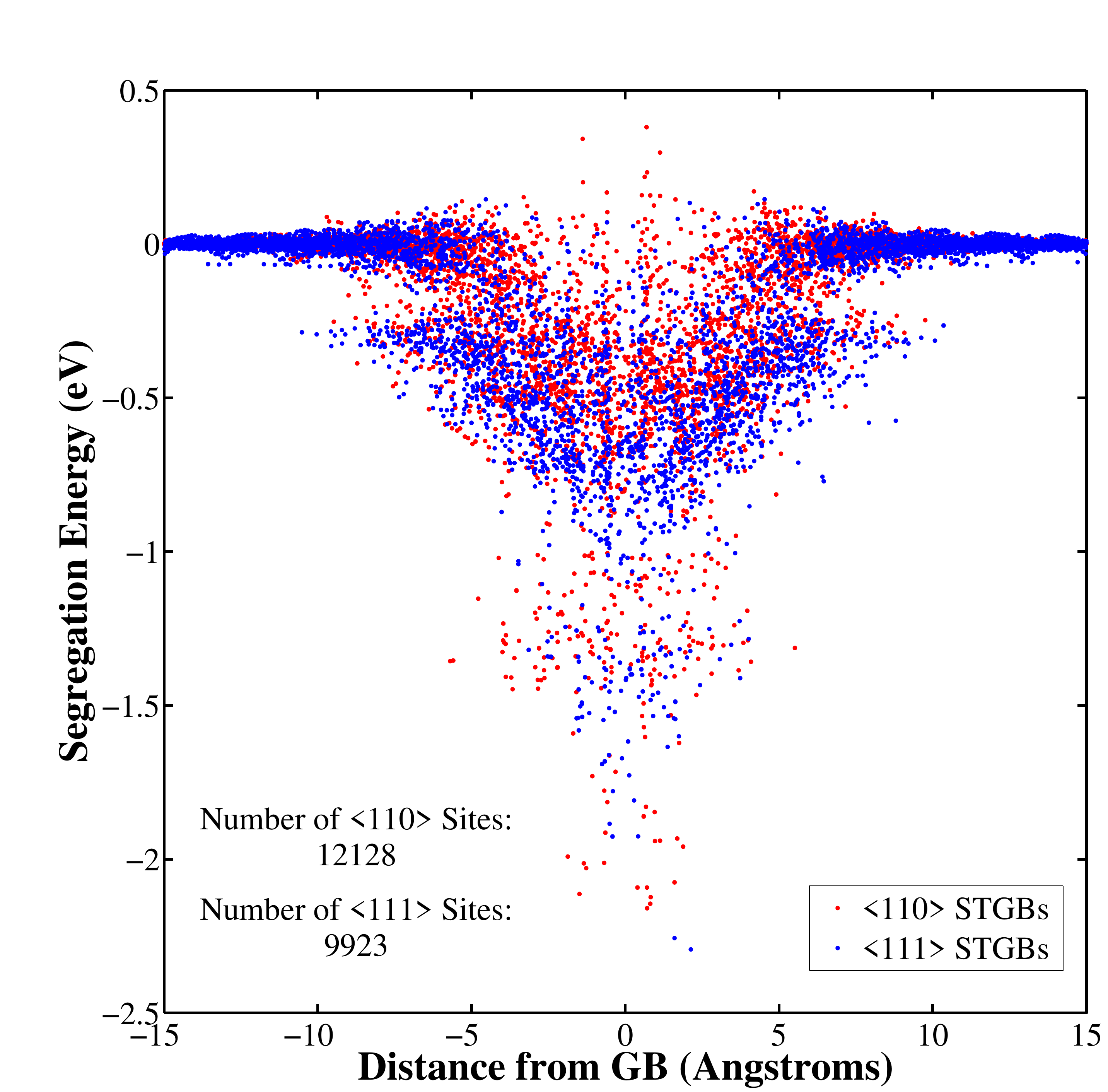}
\caption{ \label{fig:fig6} Distribution of segregation energies as a function of distance from the grain boundary for 50 \dirf110 and 25 \dirf111 STGBs.  Most grain boundary sites within 8 \AA\ have negative segregation energies that decrease with decreasing distance to the grain boundary center.}
\end{figure}

The same process used for the analysis of \dirf100 data in Figures \ref{fig:fig3}-\ref{fig:fig5} has been repeated for the data of \dirf110 and \dirf111 STGB simulations.  The distribution of segregation energies as a function of distance from the GB for all 50 \dirf110 and 25 \dirf111 STGBs is shown in Figure \ref{fig:fig6}.  This distribution is similar to that of the \dirf100 STGBs shown in Figure \ref{fig:fig4}b. However, the minimum segregation energies are much lower than that of \dirf100 STGBs and there are fewer sites with segregation energy higher than that of the bulk. 

\begin{figure}[t!]
  \centering
	\begin{tabular}{cc}
	\includegraphics[width=2.9in,angle=0]{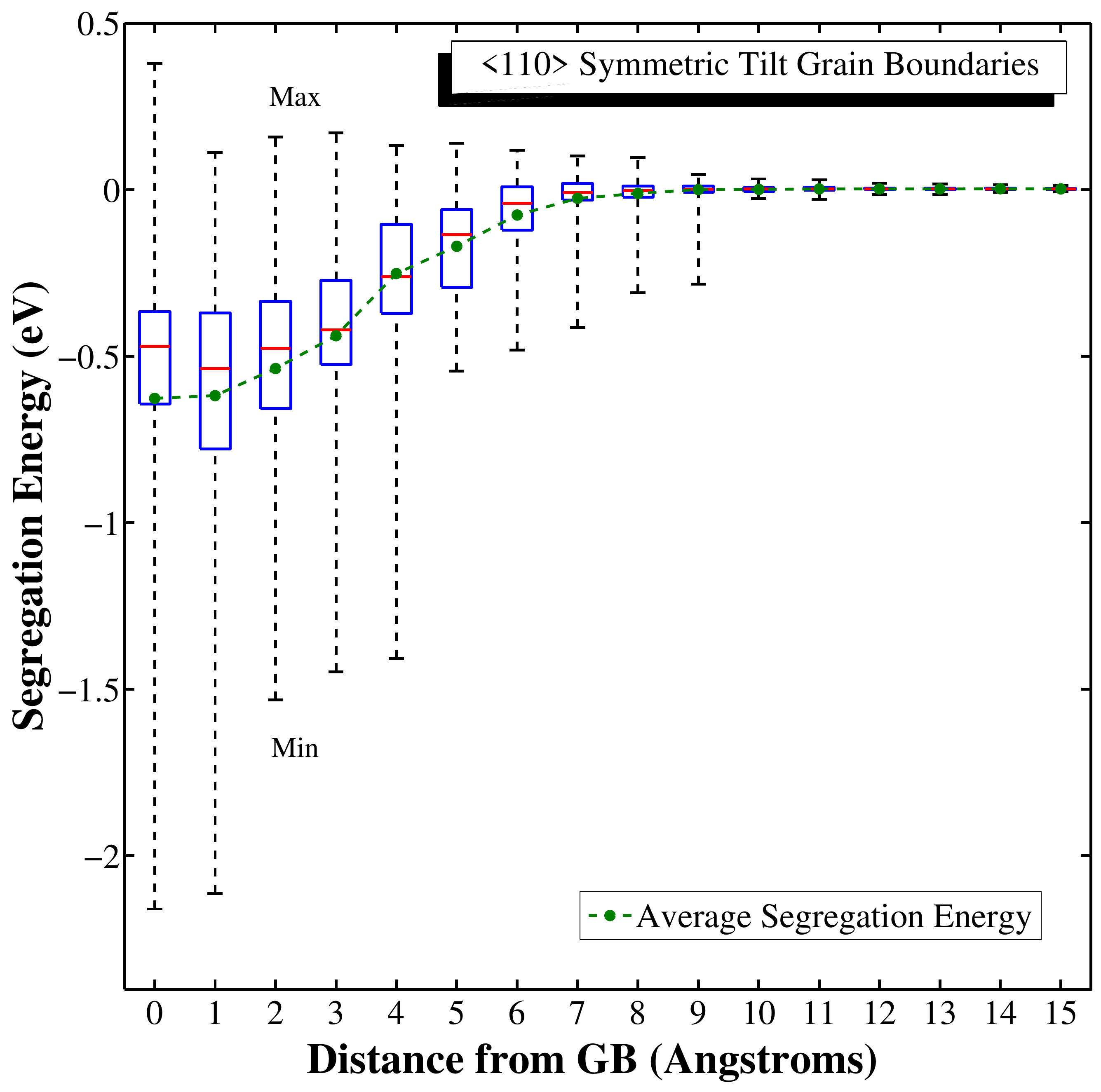} &
	\includegraphics[width=2.9in,angle=0]{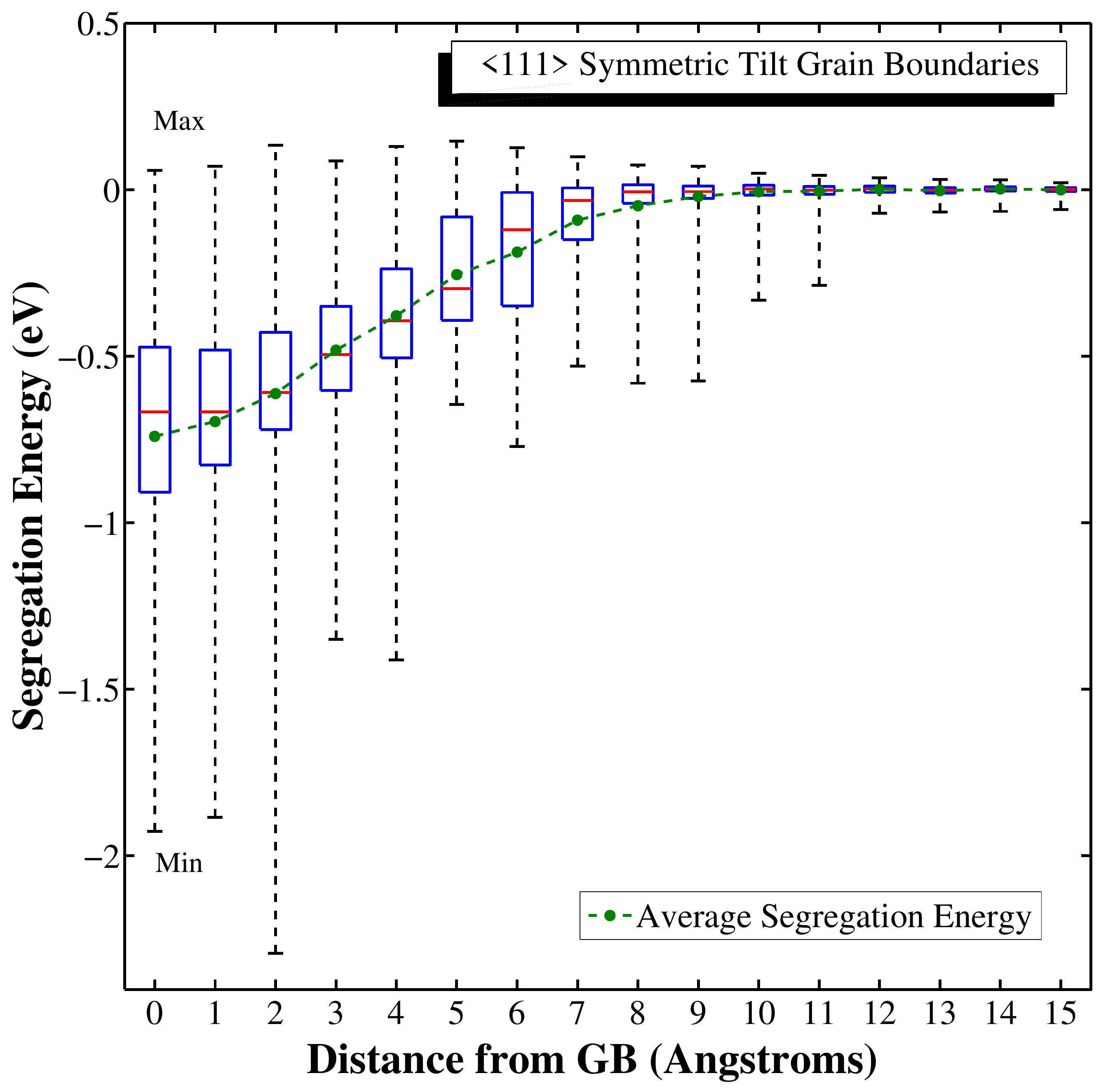} \\
	\textbf{(a) }& \textbf{(b)}
	\end{tabular}
\caption{ \label{fig:fig7} Boxplots of segregation energy as a function of distance from the grain boundary for (a) \dirf110 and (b) \dirf111 STGBs. As in Fig.~\ref{fig:fig5}, the data is divided into 1 \AA\ bins and a boxplot is made for each bin. The red lines are medians, the blue box ends are the first and third quartiles, and the black whisker ends are minimum and maximum values. The mean segregation energy is plotted in green.}
\end{figure}

A statistical representation of the data in Figure \ref{fig:fig6} is shown in Figure \ref{fig:fig7}. Similar to Fig.~\ref{fig:fig5}b, the data has been binned into 1 \AA\ bins and the median, quartiles, minimum, and maximum values of the segregation energies contained within each bin are shown within the boxplots. The mean segregation energy plots trend similarly to that in Figure \ref{fig:fig5}, though they display initially lower values close to the GB. The minimum energy whiskers again show favorable C segregation sites in most bins: up to 9 \AA\ for \dirf110 STGBs and up to 11 \AA\ for \dirf111 STGBs. 

\subsection{Statistical Characterization of Segregation Energies}

\begin{figure}[t!]
  \centering
	\begin{tabular}{c}
	\includegraphics[width=5in,angle=0]{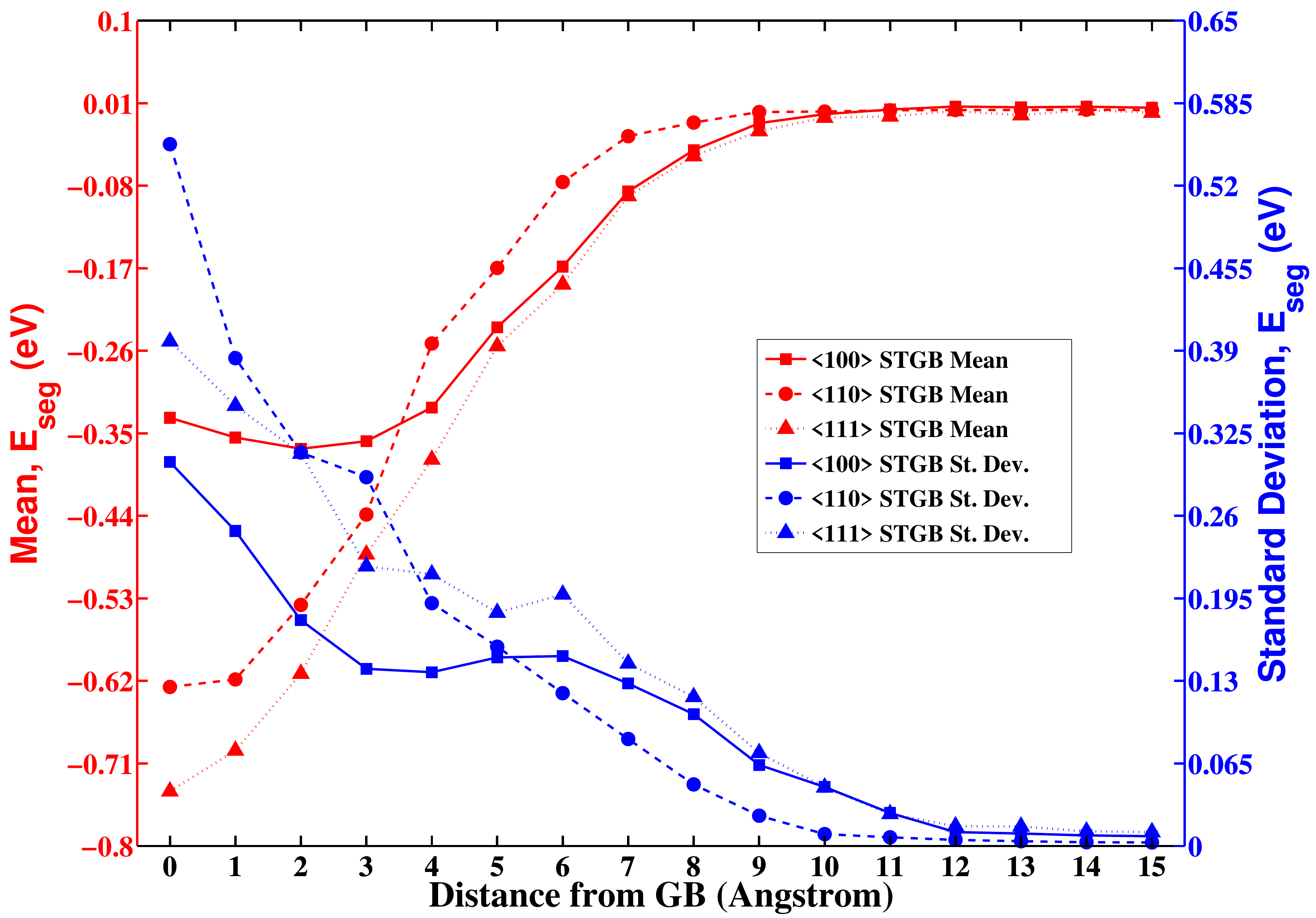} \\
	\textbf{(a) } \\
	\includegraphics[width=5in,angle=0]{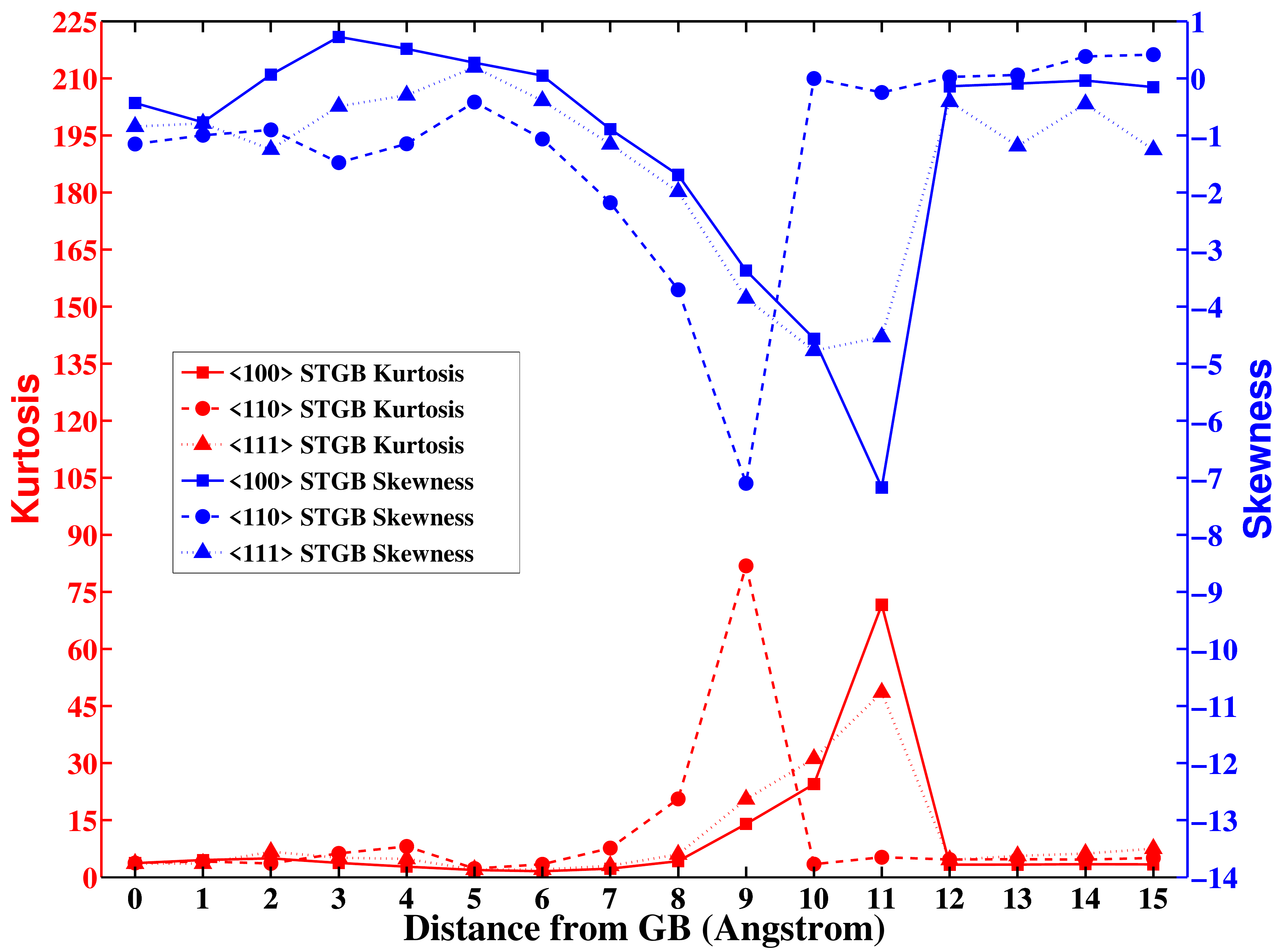} \\
	\textbf{(b)}
	\end{tabular}
\caption{ \label{fig:fig8} Statistical data for binned segregation energies from Figures 5 and 7: (a) mean and standard deviation and (b) kurtosis and skewness.}
\end{figure}

Ideally, it would be advantageous to be able to analytically describe the evolution of the segregation energies as a function of distance from the GB.  Figure \ref{fig:fig8} provides further statistical data for the binned distributions of segregation energies at given distances from the GB.  The mean and standard deviation of the segregation energy distributions are plotted in Fig.~\ref{fig:fig8}a.  For each of the GB systems, the mean values trend similarly between 5 and 15 \AA, with segregation energy decreasing with increasing distance from the GB.  However, within 5 \AA, the mean segregation energy for the \dirf100 STGB system ($E_{seg}=-0.33$ eV) is significantly higher in magnitude than the \dirf110 and \dirf111 STGB systems ($E_{seg}=-0.63$ eV and $E_{seg}=-0.74$ eV, respectively). Also plotted in Figure \ref{fig:fig8}a is the standard deviation of the distributions, which steadily decreases toward zero as distance from the GB increases. The decrease is primarily due to the increasing number of sites with bulk energy values at distances far from the boundary.  For normal distributions, the mean and standard deviation would be appropriate statistical descriptors to capture the segregation energies.  However, the boxplots in Figs.~\ref{fig:fig5} and \ref{fig:fig7} clearly show that the distributions are asymmetric to some degree and may have some extreme values or outliers.

To quantify the asymmetric distributions, the kurtosis and skewness of the distributions are plotted in Figure \ref{fig:fig8}b. The kurtosis is a measure of how heavily the variance of the distribution is affected by extreme deviations, or outliers.  Skewness is a measure of the asymmetry of a distribution and denotes in what direction a distribution possesses a longer tail of values. Figure \ref{fig:fig8}b shows that the kurtosis is relatively low for most bins but becomes very large for some GB systems at approximately 8 to 11 \AA.  This is due to most of the segregation energies approaching bulk values (as viewed by the small box, or interquartile range) except for a few negative extreme values.  Interestingly, in this range, increasing kurtosis correlates with decreasing skewness, which is negative for all but a few bins over all three GB systems.  The skewness indicates that the majority of segregation energy distributions possess longer tails of negative energies; the kurtosis indicates when these tails are typically the result of extreme deviations.  Since the magnitudes of these measures become very large at distances far from the GB, a great majority of these sites have the bulk value of segregation energy.  These four statistical parameters (mean, standard deviation, skewness, and kurtosis) can be used to better approximate asymmetric segregation energy distributions, e.g., using the Pearson system of distributions.

\section{\label{sec:sec4}Discussion}

Carbon was inserted into prior Fe sites at the GB to examine the influence of a wide range of GB structures on the distribution and magnitude of the segregation energies within the GB region.  However, in the perfect single crystal $\alpha$-Fe lattice, it is known that C occupies octahedral sites.  Hence, to accommodate the difference in energy between an octahedral C atom (and vacancy) in the bulk lattice and this same C atom occupying a substitutional site at the GB, the segregation energy associated for substitutional site $\alpha$ can be modified, i.e.,

\begin{equation}
  \label{eq:eq3}
	E_{seg^*}^{C^\alpha}=\left(E_{tot}^{GB,C^\alpha_{sub}}- E_{tot}^{GB}\right)-\left(E_{bulk}^{Fe,C_{oct}}- E_{bulk}^{Fe}\right)+\left(E_{bulk}^{Fe,vac}- E_{bulk}^{Fe}\right)
\end{equation}

\noindent where the Fe cohesive energy $E_{c}^{Fe}$ is added to the lefthand term in parenthesis to account for using the bulk energy from C in an octahedral site $E_{bulk}^{Fe,C_{oct}}$ in the righthand term.  Since the difference between Eq.~\ref{eq:eq2} and Eq.~\ref{eq:eq3} is constant, this can be calculated.  In Eq.~\ref{eq:eq2}, $(E_{bulk}^{Fe,C_{sub}}- E_{bulk}^{Fe})=0.391$ eV.  In Eq.~\ref{eq:eq3}, $(E_{bulk}^{Fe,C_{oct}}- E_{bulk}^{Fe})=-6.273$ eV (solvation energy) and $(E_{bulk}^{Fe,vac}- E_{bulk}^{Fe})=5.734$ eV.  The difference between Eq.~\ref{eq:eq2} and Eq.~\ref{eq:eq3} is $0.930$ eV.  In other words, this energy ($0.930$ eV) must be added to the preceding analysis to compare the energetic favorability of C in a substitutional site with that of an octahedral site in the bulk lattice.  Interestingly, there are a significant number of GB substitutional sites that represent a lower energy configuration for a vacancy and interstitial C in the bulk lattice ($E_{seg} < 0.930$ eV).  Specifically, the interaction length scale where this process is energetically favorable is approximately 3 \AA\ from the GB center (total width of $\approx{6}$ \AA) with the largest percentage of favorable sites occurring at the GB center: 7.2\% (within 0.5 \AA), 5.1\% (0.5-1.5 \AA), 1.1\% (1.5-2.5 \AA), 0.0\% (2.5-3.5 \AA).  Additionally, the statistical representation of the segregation energy distributions can be used to rapidly quantify the probability of lower energy sites within the GB region as a function of distance from the GB plane.  

\subsection{Substitutional versus Interstitial Carbon in Grain Boundary Region}

Interstitial sites at the GB can be assessed using a similar methodology.  There are an infinite number of potential interstitial starting positions that could be chosen since there is not a set lattice for GB interstitial sites as with substitutional sites.  The methodology chosen for selecting interstitial sites was based on the Voronoi tesselation method.  The atom positions within each GB simulation cell was used in tandem with a Voronoi tesselation of the cell to generate a list of potential starting positions for interstitial sites.  In a three-dimensional space, a Voronoi tesselation divides the space into a set of space-filling polyhedra that have the following properties: (1) any point located within a polyhedron is closest to only one atom, (2) any point on a polyhedron face is equal distance to two atoms, (3) any point on a line connecting two polyhedron faces is equal distance to three atoms, and (4) any point located on a polyhedron vertice is equal distance to four atoms.  We have chosen to use the vertices of the Voronoi tesselation to populate the set of potential interstitial sites.  In the perfect bcc lattice, the polyhedron is a truncated octahedron with 14 faces (8 regular hexagonal and 6 square), 36 edges, and 24 vertices; the Voronoi vertices are located at interstitial tetrahedral sites (i.e., equal distance to four Fe atoms).  At the GB, however, the polyhedron takes on different shapes and the Voronoi vertices constitute sites that are equal distance to four atoms, which could potentially be located at the center of GB free volume regions.  While a perturbation of this technique could be used to identify interstitial \textit{octahedral} sites in the bulk lattice\footnote{For instance, some perturbations of the present technique might be to use the midpoints of the polyhedra lines connecting the faces or use the centers of polyhedra faces.  In particular, the center of the $\left\{100\right\}$ faces formed by the four tetrahedral sites would be in exactly the minimum energy octahedral site (0,0.5$a_0$,0.5$a_0$).  However, this technique would also include the center of the $\left\{111\right\}$ faces (0.25$a_0$,0.25$a_0$,0.25$a_0$), which turns out to be a high energy interstitial position for this potential.}, the present technique is deemed sufficient for identifying potential interstitial sites in the GB.

The following Voronoi-based methodology was applied to 50 \dirf100 STGBs ($>60,000$ sites).  First, the distance that each C was displaced during minimization was calculated to examine how far each interstitial C moved from its initial site placement.  This analysis showed that most C atoms (${94.5}\%$) were displaced $<0.2$ \AA\ during the energy minimization technique, indicating that C initially placed in the tetrahedral sites tends to find a local minimum in energy and does not move to a neighboring octahedral site.  Even within the GB region, where most ($>{99.7}\%$) C atoms displaced greater than $0.2$ \AA\ lie, the maximum distance that the C atom was displaced from the initial site was only $0.59$ \AA\  ($<0.25a_0$ -- the minimum distance from a tetrahedral to an octahedral site).  Hence, the initial positions for the C interstitials identifies local minimum energy configurations centered around the tetrahedral interstitial sites in the bulk lattice, and the greatest displacements occur within the GB region, as would be expected.  Again, other perturbations of locations based on a Voronoi tesselation of the simulation cell may result in finding interstitial sites with even higher segregation energies at the boundaries, but it is anticipated that the present analysis will capture the relative influence of interstitial sites segregating to the boundary. 

\begin{figure}[t!]
  \centering
	\begin{tabular}{cc}
	\includegraphics[width=2.9in,angle=0]{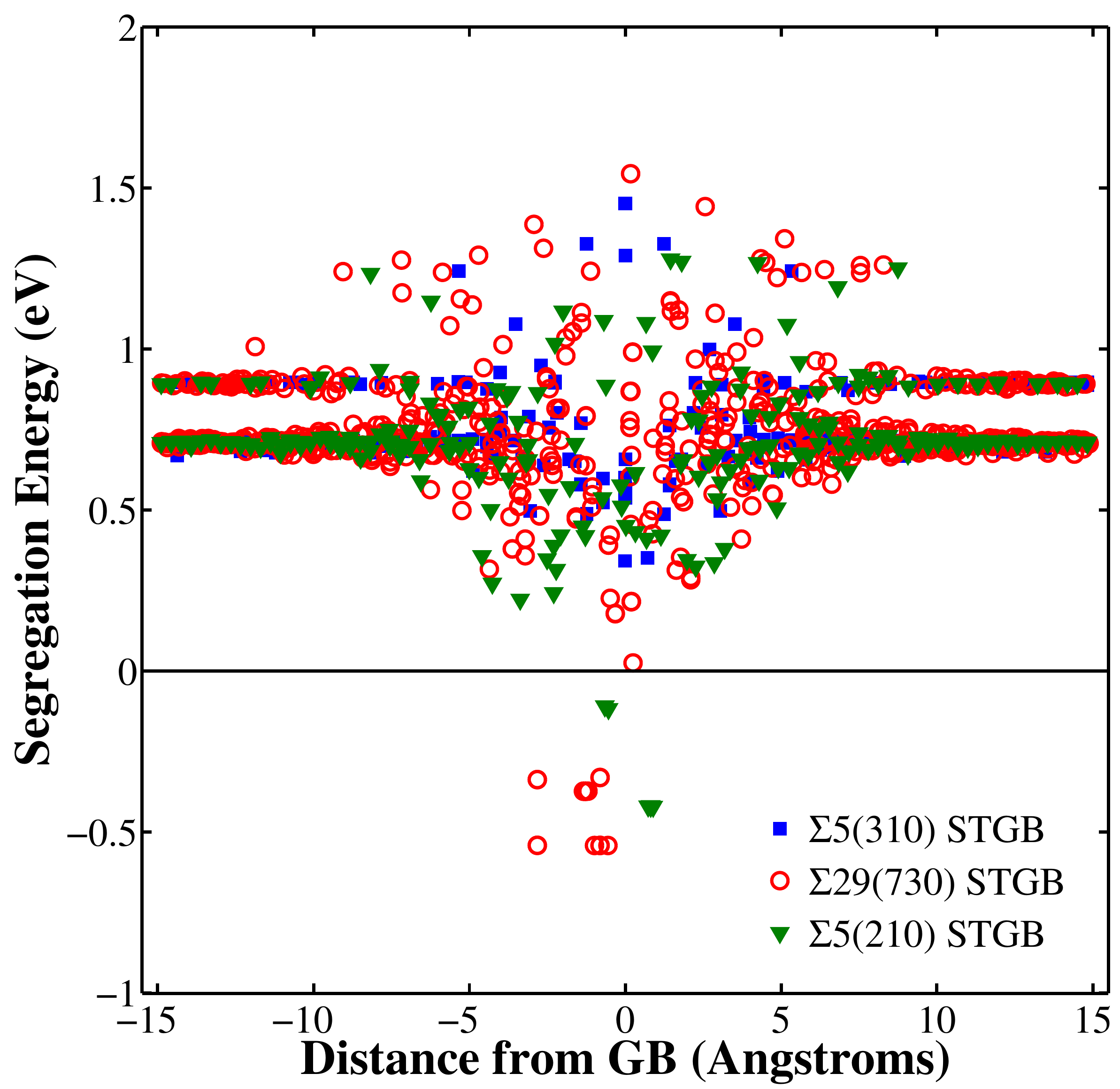} &
	\includegraphics[trim = 1.15in 0.0in 0.0in 0.5in, clip, width=2.9in,angle=0]{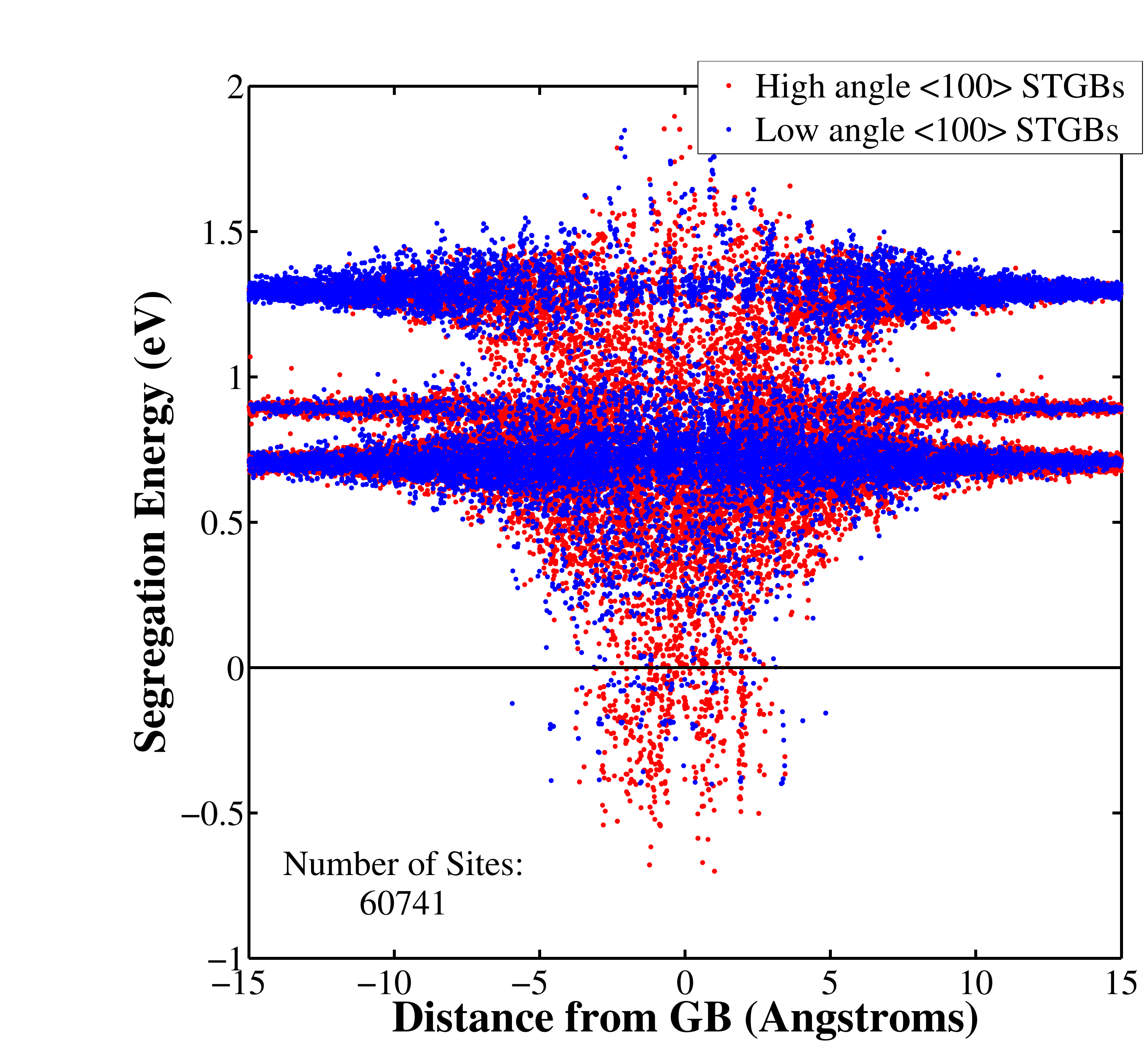} \\
	\textbf{(a) }& \textbf{(b)} \\
		\includegraphics[width=2.9in,angle=0]{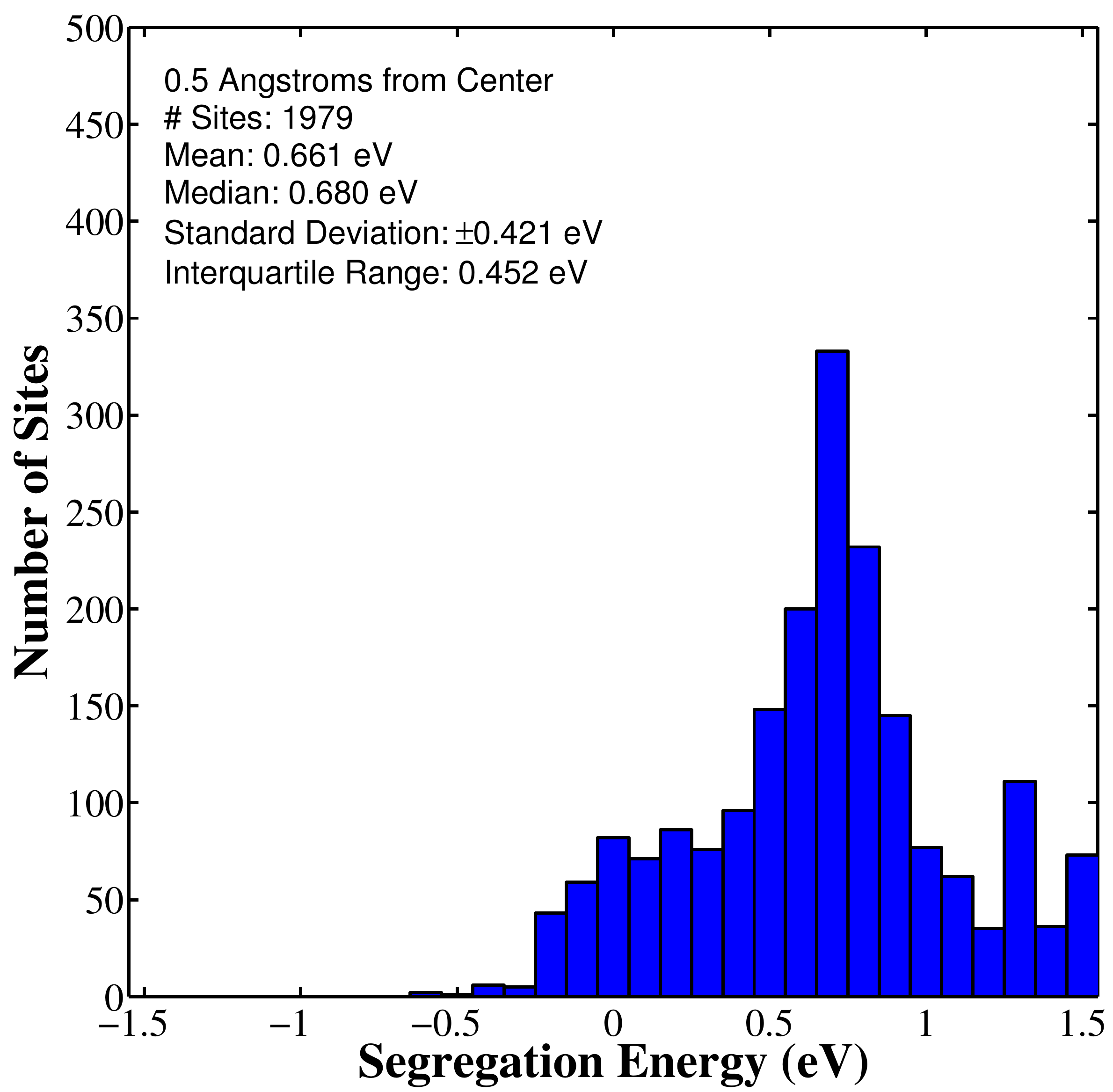} &
	\includegraphics[trim = 0.1in 0.0in 0.0in 0.1in, clip, width=2.9in,angle=0]{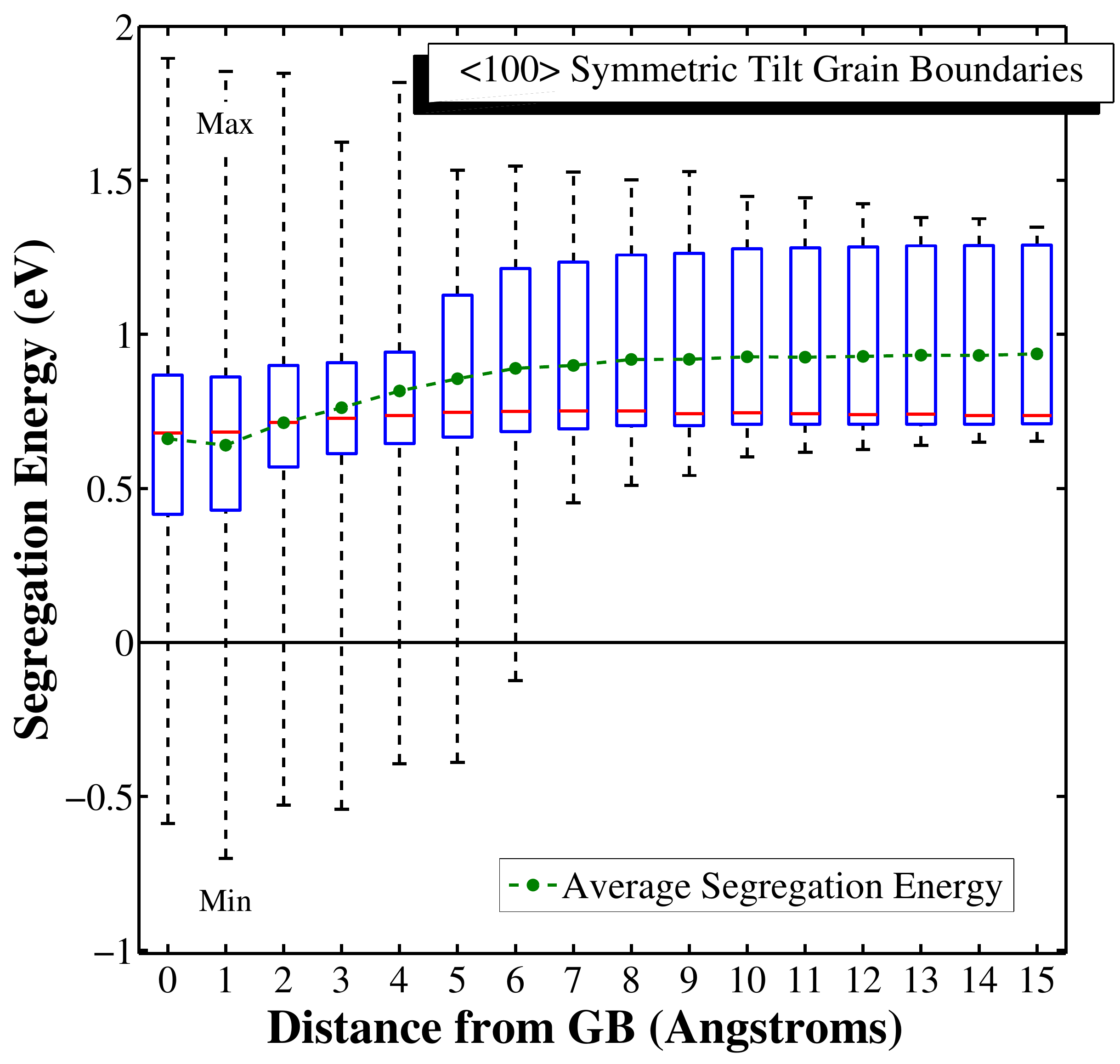} \\
	\textbf{(c) }& \textbf{(d)}
	\end{tabular}
\caption{ \label{fig:fig9} Distribution of segregation energies for interstitial sites as a function of distance from the grain boundary for (a) the $\Sigma5$\plane210, $\Sigma29$\plane730, and $\Sigma5$\plane310 grain boundaries and (b) all 50 \dirf100 STGBs, as in Fig.~\ref{fig:fig4}.  (c) The distribution of segregation energies within 0.5 \AA\ of the grain boundary center and the associated statistics.  (d) Boxplots of segregation energy as a function of distance from the grain boundary for all 50 \dirf100 STGBs, divided into 1 \AA\ bins and colored as in Fig.~\ref{fig:fig5}.}
\end{figure}

The results of inserting C atoms at interstitial sites was then analyzed in a similar manner to C placed at the substitutional sites.  In this analysis, Equation \ref{eq:eq1} is modified such that the segregation energy associated with a C atom at site $\alpha$, $E_{seg}^{C^\alpha_{int}}$ is calculated by

\begin{equation}
  \label{eq:eq4}
	E_{seg}^{C^\alpha_{int}}=\left(E_{tot}^{GB,C^\alpha_{int}}- E_{tot}^{GB}\right)-\left(E_{bulk}^{Fe,C_{oct}}- E_{bulk}^{Fe}\right)	
\end{equation}

\noindent where $E_{tot}^{GB,C^\alpha_{int}}$ is the total energy of the GB structure with an interstitial C atom.  Figure \ref{fig:fig9}(a-d) corresponds to Figs.~\ref{fig:fig4} and \ref{fig:fig5}, except that the calculated segregation energies are for interstitial C using Eq.~\ref{eq:eq4}.  There are several minimum energy states for interstitial C far away from the GB (Fig.~\ref{fig:fig9}(a)).  The energy of $\approx0.887$ eV corresponds to the tetrahedral site.  However, the present Fe-C potential also has minimum energy interstitial sites at other locations, and the Voronoi vertex technique did not locate the octahedral site for the C atom upon energy minimization (i.e., $E_{seg}^{C^\alpha_{int}}=0$ is noticeably absent at large distances).  Additional simulations varying the location of interstitial positions in a 10$a_0$x10$a_0$x10$a_0$ bcc cell with 2000 atoms show that one of the high energy sites is directly between two Fe atoms along the \dirf111 direction (0.25$a_0$,0.25$a_0$,0.25$a_0$).  In Fig.~\ref{fig:fig9}(a), the distribution of segregation energies as a function of distance is shown for the same three GB structures as in Fig.~\ref{fig:fig4}(a).  The GB region again shows both energetically favorable and unfavorable sites, with several energetically-favorable interstitial sites having $E_{seg}^{C^\alpha_{int}}$ of up to $-0.5$ eV (i.e., a binding energy of approximately $0.5$ eV).  This same trend is also evident for all 50 \dirf100 STGBs (Fig.~\ref{fig:fig9}(b)).  In fact, based on this plot, the interaction length scale ($E_{seg}^{C^\alpha_{int}}<0$) of this GB system is on the order of 10 \AA\ or less.  Binning the data from this plot, the distribution of segregation energies within 0.5 \AA\ from the GB center (Fig.~\ref{fig:fig9}(c)) show an asymmetric multimodal character with a few peaks centered about the several minimum energy interstitial states observed at large distances from the GBs.  Approximately 6.9\% of interstitial sites sampled showed a lower energy than C at an octahedral site in the bulk lattice.  The boxplots of interstitial C shows a decrease in the mean segregation energy, the interquartile range, and the minimum segregation energy starting at $\approx{5}$ \AA\ from the GB center (Fig.~\ref{fig:fig9}(d)).  Interestingly, there is a larger percentage (11.8\%) of energetically-favorable sites in the second bin (0.5-1.5 \AA\ from GB center) than in the first bin (within 0.5 \AA) and this percentage decreases with increasing distance from the boundary: 6.9\% (0.5 \AA), 11.8\% (0.5-1.5 \AA), 6.2\% (1.5-2.5 \AA), 3.0\% (2.5-3.5 \AA), 0.3\% (3.5-4.5 \AA), and 0.2\% (4.5-5.5 \AA).  There is a high degree of anisotropy in the segregation energies in each bin due to the GB character.  Moreover, in contrast to the energetic length scales for point defects in $\alpha$-Fe using the same interatomic potential, the calculated length scales of interaction between C and the GB are much lower. 

\subsection{Methodology application and extensions}

There are a number of studies that examine the influence of various alloying elements and impurities with grain boundaries.  For the case of steels, while the interaction between GB structures and C represents one example of segregation that has been experimentally observed \cite{Cow1998,Pap1971,Tak2005}, many studies also focus on which elements segregate to the GB and how these elements may interact with other alloying elements.  For instance, elements that can impact the properties of steels include, for example, phosphorus, sulfur, hydrogen, nitrogen, manganese, silicon, molybdenum, nickel, chromium, antimony, and tin \cite[e.g.,][]{Ban1978, Bri1978, Yu1980a, Yu1980b, Kam1981}.  Also, experiments have shown that there may be competitive processes between elements (e.g., C and phosphorus \cite{Cow1998}) at the GB that will also depend on processing conditions, such as equilibration temperature and quench rate.  Of course, this is just one example of the complexity of segregation in steels due to the numerous interactions of different elements with grain boundaries and the kinetics of those processes.  The present work may be modified to examine how binding behavior between two different atom species (e.g., C-P, C-vacancy) or different atom configurations (e.g., $H$ vs. $H_2$ \cite{Sol2012}) are affected by proximity to and within grain boundaries.  Understanding these interactions may aid in understanding the complexity of segregation and its subsequent impact on properties in $\alpha$-Fe and alloy steels.

This methodology makes possible the statistical representation of impurity segregation to grain boundaries while accounting for differences between GB structures. Thus, the results from this method could be used as inputs in other simulations, such as the kinetic Monte Carlo technique, mesoscale models, or analytical models.  In this sense, the information being passed is not just scalar values but is distributions of values, which can be used to analyze sensitivity and incorporate variability due to GB structure within multiscale models.  Clearly, the present application of calculating these distributions for segregation of C to grain boundaries represents one such example of this concept.  Moreover, such studies could be expanded upon to include the the effects of temperature with different solutes and solute concentrations.  For example, Rittner and Seidman \cite{Rit1997} conducted such a study for 21 \dirf110 GBs to calculate segregation free energies, entropies, and internal energies for a Ni-Pd system. Their work found a linear relation of segregation internal energies and entropies, which suggests the possibility for estimating segregation free energies from internal energies, an easier quantity to calculate. Simulations at temperature may lead to an even better prediction of segregation behavior to grain boundaries in polycrystalline materials.  

\section{\label{sec:sec5}Conclusion}

In this work, we have used molecular statics simulations to investigate the segregation energy of a single C atom to thousands of substitutional and interstitial atomic sites in 50 \dirf100, 50 \dirf110, and 25 \dirf111 STGBs.  A large number of boundaries, including general low and high angle GBs, were used in order to account for the variability in GB degrees of freedom observed in experimental polycrystalline materials. We can draw the following conclusions based upon our results:

\begin{enumerate}
	\item A methodology for calculating and analyzing the segregation energies of thousands of sites within a large number of grain boundaries with molecular statics simulations has been developed.  This method samples different boundaries from a grain boundary database and calculates the segregation energy for every grain boundary site to acquire segregation statistics.  As a first example, C segregation to $\alpha$-Fe boundaries was examined.  Both substitutional sites were sampled as well as potential interstitial sites, using the polyhedra vertices calculated using a Voronoi tesselation of the three-dimensional $\alpha$-Fe coordinates.  Such a methodology is warranted given that we found a large degree of anisotropy in sites and segregation energies due to varying grain boundary structure.
	\item The local structure within the grain boundary affects the segregation energy.  As an example, the \dirf100 symmetric tilt system is shown where the two $\Sigma5$ grain boundaries are both cusps in the energy relationship (Fig.~\ref{fig:fig1}) and contain the favored structural units of this system (Fig.~\ref{fig:fig2}).  However, boundaries of intermediate misorientations (e.g., the $\Sigma29$ boundary) - which contain combinations of the same structural units - do not necessarily have the same segregation energy distributions as the $\Sigma5$ boundaries (much lower, see Fig.~\ref{fig:fig3}).
	\item For the substitutional C atom case, we found that it is energetically favorable for interstitial octahedral C and a vacancy in the lattice to combine within the grain boundary at a substitutional site.  While this process is highly unfavorable in the lattice, there is a region that extends $\approx{3}$ \AA\ from the grain boundary center where there are favorable substitutional sites for the grain boundaries sampled.  The largest percentage of favorable sites are directly at the grain boundary center. 
	\item For the interstitial C case, we found that it is energetically favorable for a C atom at an octahedral site in the lattice to segregate to the grain boundary with a maximum binding energy of $\approx{0.5}$ eV.  The interaction length scale of the grain boundary with octahedral C is $\approx{5}$ \AA\ from the grain boundary center with the largest percentage of favorable sites located within the bin just outside of the grain boundary center (0.5 \AA\ to 1.5 \AA\ from the grain boundary center).
	\item To quantify the segregation energy distributions as a function of distance from the grain boundary, the energies were separated into 1 \AA\ bins and characterized using several statistical descriptors: quartile values, median, mean, and extreme values (see Figs.~\ref{fig:fig5} and \ref{fig:fig7}).  The grain boundary atomic sites have asymmetric distributions of segregation energy with some extreme values that extend over 10 \AA\ from the grain boundary.  Furthermore, close to the grain boundary, the majority of these distributions are negatively skewed, indicating longer tails of negative segregation energies.   An analytical model informed by these calculations whereby the segregation energy distribution as a function of distance is captured using four statistical parameters (mean, standard deviation, kurtosis, skewness - see Fig.~\ref{fig:fig8}) is hypothesized for upscaling to higher scale models, i.e., parameters necessary for a Pearson system of distributions.
\end{enumerate}

The significance of this research is not just the calculations of the energetics of C segregation in a specific class of grain boundaries in $\alpha$-Fe, but also the development of a methodology capable of ascertaining segregation energies over a wide range of grain boundary character typical of that observed in polycrystalline materials.

\section*{Acknowledgments}

MAT would like to acknowledge funding under the U.S.~Department of Energy and the National Energy Technology Laboratory under Award Number DE-FC26-02OR22910 and the U.S.~Department of Energy's Nuclear Energy Advanced Modeling and Simulation (NEAMS) Program at the Pacific Northwest National Laboratory, which is operated by Battelle under Contract No.~DE-AC05-76RL01830. KNS would like to acknowledge the support by the Office of Naval Research under contract No.~N000141110793.  This report was prepared as an account of work sponsored by an agency of the United States Government. Neither the United States Government nor any agency thereof, nor any of their employees, makes any warranty, express or implied, or assumes any legal liability or responsibility for the accuracy, completeness, or usefulness of any information, apparatus, product, or process disclosed, or represents that its use would not infringe privately owned rights. Reference herein to any specific commercial product, process, or service by trade name, trademark, manufacturer, or otherwise does not necessarily constitute or imply its endorsement, recommendation, or favoring by the United States Government or any agency thereof. The views and opinions of authors expressed herein do not necessarily state or reflect those of the United States Government or any agency thereof. Such support does not constitute an endorsement by the Department of Energy of the work or the views expressed herein.  

\def\newblock{\hskip .11em plus .33em minus .07em}

\bibliographystyle{unsrt}

\end{document}